\documentclass[aps,prl, superscriptaddress,groupedaddress]{revtex4}                           

\usepackage{graphicx}
\usepackage{amsmath}
\usepackage{amssymb}
\usepackage[amssymb]{SIunits}
\usepackage{color}
\usepackage[usenames,dvipsnames]{xcolor}
\usepackage{ulem}
\usepackage{cancel}
\usepackage{natbib}
\usepackage[T1]{fontenc}
\usepackage{bm}
\usepackage{graphicx}
\usepackage{epstopdf, epsfig}
\usepackage{color}
\usepackage{soul}
\usepackage{subfigure}
\usepackage{amsmath}
\usepackage{lineno}
\usepackage{setspace}
\usepackage{longtable}
\makeatletter

\begin{document}
\setstretch{1.2} 
\title{Effects of radius ratio on annular centrifugal Rayleigh-B\'{e}nard convection}

\author{Dongpu Wang}
\affiliation{Center for Combustion Energy, Key Laboratory for Thermal Science and Power Engineering of MoE, and Department of Energy and Power Engineering, Tsinghua University, 100084 Beijing, China.}

\author{Hechuan Jiang}
\thanks{jhcthu@foxmail.com}
\affiliation{Center for Combustion Energy, Key Laboratory for Thermal Science and Power Engineering of MoE, and Department of Energy and Power Engineering, Tsinghua University, 100084 Beijing, China.}

\author{Shuang Liu}
\affiliation{Center for Combustion Energy, Key Laboratory for Thermal Science and Power Engineering of MoE, and Department of Energy and Power Engineering, Tsinghua University, 100084 Beijing, China.}

\author{Xiaojue Zhu}
\affiliation{Max Planck Institute for Solar System Research, Justus-von-Liebig-Weg 3, 37077 G{\"o}ttingen, Germany}

\author{Chao Sun}
\affiliation{Center for Combustion Energy, Key Laboratory for Thermal Science and Power Engineering of MoE, and Department of Energy and Power Engineering, Tsinghua University, 100084 Beijing, China.}
\affiliation{Department of Engineering Mechanics, School of Aerospace Engineering, Tsinghua University, 100084 Beijing, China}

\date{\today}

\begin{abstract} 
{\bf We report on a three-dimensional direct numerical simulation study of flow structure and heat transport in the annular centrifugal Rayleigh-B\'{e}nard convection (ACRBC) system, with cold inner and hot outer cylinders corotating axially, for the Rayleigh number range Ra $ \in [{10^6},{10^8}]$ and radius ratio range $\eta  = {R_i}/{R_o} \in [0.3,0.9]$. This study focuses on the dependence of flow properties on the radius ratio $\eta$.  The temperature and velocity fields reveal that different curvatures of the inner and outer cylinders of the ACRBC system lead to asymmetric movements of hot and cold plumes under the action of Coriolis force, resulting in the formation of zonal flow. The physical mechanism of zonal flow is verified by the dependence of the drift frequency of the large-scale circulation rolls and the space- and time-averaged azimuthal velocity on $\eta$. We find that the larger $\eta$ is, the weaker the zonal flow becomes. We show that the heat transport efficiency increases with $\eta$. It is also found that the bulk temperature deviates from the arithmetic mean temperature and the deviation increases as $\eta$ decreases. This effect can be explained by a simple model that accounts for the curvature effects and the radially-dependent centrifugal force in ACRBC. }
\end{abstract}

\maketitle
\section{Introduction}
Turbulent convection is ubiquitous in nature and in many industrial processes. Examples include the convective flows in the Earth’s mantle \citep{mantle} and outer core \citep{outer_core}, in the atmospheric motion \citep{atmosphere1,atmosphere2}, in the ocean \citep{ocean}, and in rotational machines \citep{engineering}. Many of these convection phenomena occur under rapid rotation of the system \citep{Hide1975,Bohn1995}. Rayleigh-B\'{e}nard convection (RBC), a fluid layer heated from below and cooled from above, is a classical and idealized paradigm for the study of thermally driven turbulent flows \citep{Review1,Review2,Review3,Review4,Acta1,Acta2,chen_wang_xi_2020}. The main issues for thermal turbulence studies include the dynamics of turbulent structures and the scaling relation between the heat transport, in the dimensionless form, Nusselt number Nu, and thermally driven force, in the dimensionless form, Rayleigh number Ra. Recently a novel system similar to classical RBC, Annular Centrifugal RBC (ACRBC) system with cold inner and hot outer cylinders corotating axially, has been proposed \citep{Kang2019PRF,JiangSciAdv,rouhi2021}. By exploiting strong centrifugal force through rapid rotation, the intensity of the thermal driving can be significantly enhanced. 

In ACRBC, Jiang et al. \cite{JiangSciAdv} found that the convective rolls revolve around the rotating center in prograde direction, signifying the emergence of zonal flow, which may be related to the effects of Coriolis force and the different curvatures of the two cylinders. In astrophysical and geophysical studies, by using a rotating cylindrical annulus with conical end surfaces, Busse and his collaborators \citep{Busse1,Busse2,Busse3,Busse4} also observed zonal flow phenomenon. According to the topographic-$\beta$ approximation \citep{Yano_topographic_approx}, the strength and direction of the zonal flow in their system depend on the radial gradient of the axial fluid column height. By using a spherical shell model with a radius ratio of 0.9, Heimpel et al. \cite{Jupiter/Nature} found that zonal flow in the equatorial latitude of Jupiter is prograde with respect to the planet, which is consistent with the actually observed results of Jupiter \citep{Jupiter/picture}, but the width of equatorial zonal flow does not coincide with the predictions of Rhines scale \citep{rhines_1975}. The physical mechanism of zonal flow in ACRBC and how it is affected by the curvatures of the two cylinders \citep{JiangSciAdv} deserve further study.

In classical turbulent RBC, the effects of aspect ratio $\Gamma$ on the Nusselt number Nu have been extensively studied \citep{sun_ren_song_xia_2005,Review1,Poel2011PRE,Huang2013PRL,huang_xia_2016} and it is found that heat transport efficiency has a great relevance on $\Gamma$ when it is smaller than 1 \citep{Poel2011PRE}. Huang et al. \cite{Huang2013PRL,huang_xia_2016} investigated the effects of lateral confinement on heat transport in quasi-2D turbulent RBC and found that narrow lateral width of the convection cell induces the increase of the heat transfer efficiency. In high-Reynolds number Taylor-Couette (TC) turbulence, Grossmann et al. \cite{Sun2016Annu} analyzed multiple sets of data from the previous experiments and direct numerical simulations, and found that in the range of radius ratio 0.5-0.909, as the radius ratio increases, the amplitude of TC Nusselt number $\rm {N{u_\omega }}$ (the dimensionless angular velocity flux) first increases, and then saturates when the radius ratio is greater than 0.7, indicating that larger radius ratio can achieve higher momentum transport efficiency. Therefore, it is of vital importance to study the effects of geometry on the heat transport efficiency of the ACRBC system, and it can also give insights for the design of rotating machinery \citep{design1,design2,design3}.

We notice that Pitz et al. \cite{pitz_marxen_chew_2017}, Kang et al. \cite{Kang2019PRF} have studied effects of the radius ratio on the centrifugal buoyancy driven flow, but they mainly focused on the dependence of the critical Rayleigh number $\rm {R{a_c}}$ and the critical wavenumber ${\omega _c}$ of convection onset through linear stability analysis and numerical simulation. For higher Rayleigh number Ra in the turbulent regime, to the best of our knowledge few attentions have been paid to systematically studying the effects of radius ratio. To fill this gap, we present a systematic investigation of the dynamics of zonal flow and heat transfer properties in the turbulent regime of the ACRBC system with radius ratio from 0.3 to 0.9 by means of high-resolution three-dimensional (3D) direct numerical simulation (DNS). Will the geometric effects in ACRBC  be similar to that in RBC or in TC? Answering this question is the major objective in this study. 

The remainder of this manuscript is organized as follows. In $\S$ 2, we give a brief description of the governing equations and the numerical model.  The results are presented and analyzed in $\S$ 3, which is divided into three parts, $\S$ 3.1 describes the dynamics of zonal flow in ACRBC and explains the physical mechanism of it. In $\S$ 3.2, we show the dependence of heat transfer on radius ratio, and discuss the physical reasons. Remarkably asymmetric mean temperature fields are found in ACRBC and predicted by a theoretical model, which are discussed in $\S$ 3.3. Finally, we summarize our findings in $\S$ 4.
\section{Numerical settings}
\subsection{Flow setup}
\begin{figure}
	\centerline{\includegraphics[width=0.99\linewidth]{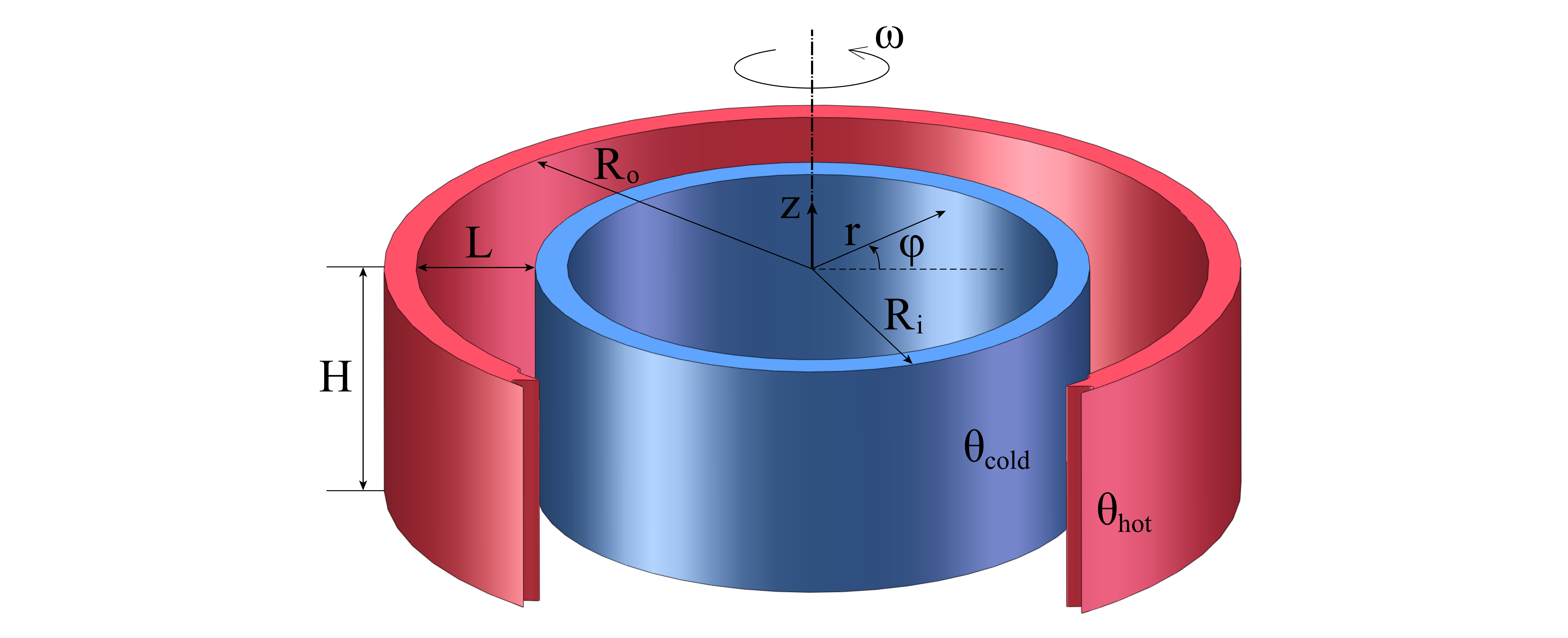}}
	\caption{
		Schematic diagram of the flow configuration. $\omega$ is the angular velocity of the system. All cases in this paper set the angular velocity unit vector $\boldsymbol{\widehat \omega}$ = +1, that is, the system rotates counterclockwise.  ${R_o}$, ${R_i}$, $H$ and $L$ are the inner radius of the outer cylinder, the outer radius of the inner cylinder, the height of the cylindrical annulus, and the gap width between the two cylinders, respectively. ${\theta _{hot}}$ and ${\theta _{cold}}$ denote the temperature of the outer and inner walls.
	}
	\label{Fig1}
\end{figure}

We consider a fluid bounded by cold inner and hot outer cylinders, which corotate axially as shown in figure \ref{Fig1}. The governing equations are derived from Navier-Stokes equations under the Boussinesq approximation in a rotating reference frame, which are expressed as:
\begin{small}
	\begin{align}
	\nabla \cdot {\bf u} &= 0 \ ,\\
	\frac{\partial \theta}{\partial t} +{\bf u}\cdot \nabla \theta &= \frac{1}{\sqrt{RaPr}}\nabla^2 \theta \ ,\\
	\frac{\partial \bf{u}}{\partial t} +{\bf u}\cdot \nabla {\bf u} &= -\nabla p+\text{Ro}^{-1} \hat{\bm{\omega}} \times {\bf u} \nonumber \\ 
	&\phantom{=} +\sqrt{\frac{Pr}{Ra}}\nabla^2 {\bf u} -\theta \frac{2(1-\eta)}{(1+\eta)}\bf r \ ,
	\end{align}
\end{small}
\noindent where $\boldsymbol{\widehat \omega}$ is the unit vector pointing in the direction of the angular velocity, $\boldsymbol{u}$ is the velocity vector normalized by the free-fall velocity $U \equiv \sqrt {{\omega ^2}\frac{{({R_o} + {R_i})}}{2}\alpha \Delta L} $, $t$ is the dimensionless time normalized by $L/U$, and $\theta$ is the temperature normalized by $\Delta$. Here, as defined in figure \ref{Fig1}, $\omega$ denotes the angular velocity of the system.  ${R_o}$ and ${R_i}$ are the radius of the outer and inner cylinders, respectively. $\alpha$,$\nu$,$\kappa$ are isobaric thermal expansion coefficient, kinematic viscosity and thermal diffusivity of the fluid, respectively. $\Delta$ and ${L}$ are the temperature difference ($\Delta \equiv {\theta _{hot}} - {\theta _{cold}}$) and the gap ($L \equiv  {R_o} - {R_i}$) between the two cylinders. The coordinate system  $\phi$, $z$, $r$ refer to the streamwise (azimuthal), spanwise (axial) and wall-normal (radial) directions.

The above dimensionless governing equations reveal that ACRBC is mainly controlled by four dimensionless parameters. Similar to classical RBC, Rayleigh number defined as
\begin{small}
	\begin{align}
	{\rm Ra} &= \frac{1}{2}{\omega ^2}({R_o} + {R_i})\alpha \Delta {L^3}/(\nu \kappa ) \ ,
	\end{align}
\end{small}
\noindent and Prandtl number $\rm {Pr} = $ $ \nu /\kappa$ characterize buoyancy-driven strength and physical properties of the convecting fluid.
Two additional control parameters are $\rm {R{o^{ - 1}}} = $ $\omega L/U$ and $\eta  = {R_i}/{R_o}$, which measure Coriolis force effects and geometric properties. The key response parameter is the Nusselt number given by
\begin{small}
	\begin{align}
	{\rm Nu} &= \frac{J}{{{J_{con}}}} = \frac{{{{\left\langle {{v_r}\theta} \right\rangle }_{\phi ,z,t}} - \kappa \frac{\partial }{{\partial r}}{{\left\langle \theta \right\rangle }_{\phi ,z,t}}}}{{\kappa \Delta {{(r \cdot ln(\eta ))}^{ - 1}}}} \ ,
	\end{align}
\end{small}
\noindent where $J$, ${{J_{con}}}$, ${v_r}$ and $\theta$ denote the total heat flux, the heat flux through pure thermal conduction, the radial velocity and temperature of a certain point, respectively. ${\left\langle {...} \right\rangle _{\phi ,z,t}}$ denotes taking average over the $\phi z$-plane and time.
\subsection{Direct numerical simulations}
Numerical simulations are performed using an energy-conserving second-order finite-difference code, which has been described detailedly in literature \citep{VERZICCO1996402,VANDERPOEL201510,Zhu2018comput,JiangSciAdv}. Thus, here we introduce briefly and just give some main features. 
Table \ref{tab1} documented in appendix lists all the specific simulation parameters. For all cases, no-slip boundary conditions for velocity and constant temperature boundary conditions are adopted at surfaces of inner and outer cylinders. Periodic boundary conditions are imposed on $\boldsymbol{u}$ and $\theta$ in the $z$-direction. The aspect ratio ($\Gamma  = H/L$) for most cases is set  $\Gamma  = 1$, but for high Ra cases (i.e. Ra $> {10^7}$), $\Gamma$ is reduced to 0.5 even 0.25. Furthermore, for large Ra and $0.6 \le \eta  \le 0.9$, the azimuthal domain is reduced from a whole circle (${\phi _0} = 1$) to ${\phi _0} = 1/2$, $1/4$, $1/8$. The flow domain at least contains a pair of convection rolls to ensure the statistical stability. It is found that such an arrangement will not have considerable impact on Nu and zonal flow.
\begin{figure*}
	\centerline{\includegraphics[width=0.99\linewidth]{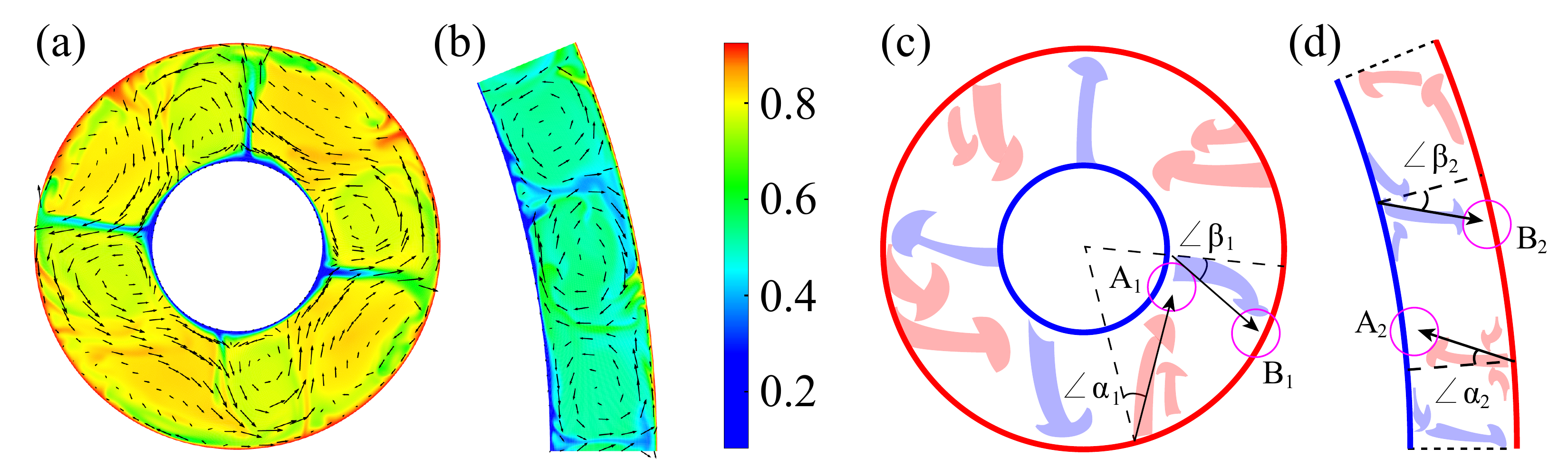}}
	\caption{
		Typical snapshots of instantaneous temperature fields, superposed by the velocity vectors for $\eta$ = 0.4 (a) and $\eta$ = 0.9 (b) at $\rm Ra = {10^7}$, $\rm R{o^{ - 1}} = 1$ and $\Pr  = 4.3$. (c,d) Corresponding sketches to (a) and (b), which show the motions of cold (blue) and hot (red) plumes.
	}
	\label{Fig2}
\end{figure*}

Adequate resolutions are ensured for all simulations and we have performed a posteriori check of spatial and temporal resolutions to guarantee to resolve all relevant scales. As shown in table \ref{tab1}, the ratio of maximum grid spacing ${\Delta _g}$ in the bulk region to the Kolmogorov scale estimated by the global criterion ${\eta _K}{\rm{ = }}\frac{{L{{\Pr }^{1/2}}}}{{[\rm{Ra(Nu - 1)}]}^{1/4}} \cdot {[\frac{{(1 + \eta )\ln (\eta )}}{{2(\eta  - 1)}}]^{1/4}}$ \citep{JiangSciAdv} is smaller than 0.7 (${\Delta _g}/{\eta _K}  < 0.7$). We have also compared ${\Delta _g}$ with the Batchelor scale ${\eta _B} = {\eta _K}{\Pr ^{ - 1/2}}$ \citep{Silano2010} for each case (not shown here) and we have ${\Delta _g}/{\eta _B} < 1.4$. Furthermore, the clipped Chebychev-type clustering grids adopted in the radial direction ensure the spatial resolution within BLs. There are at least 8 grid points inside thermal BLs and 10 grid points inside viscous BLs. We use the Courant–Friedrichs–Lewy (CFL) conditions to check temporal resolution \citep{CFL1928,VANDERPOEL201510,zhang_zhou_sun_2017}, i.e. the CFL number is smaller than 0.8 for all simulations to ensure computational stability. ${\tau _{avg}}$ is the averaging time for Nusselt number. For simulation convergence, each case is run for about 100 free-fall time units to discard initial transients, and we obtain the Nu by averaging over an additional ${\tau _{avg}} \ge 80$ and over the Nusselt numbers $\rm Nu_{in}$ and $\rm Nu_{out}$ at the inner and outer walls. One way for statistical convergence is when the difference of Nu between inner and outer walls ${\epsilon _{\rm Nu}} =$ $\rm \left| {N{u_{in}} - N{u_{out}}} \right|/Nu$ is small and acceptable \citep{ostilla2013,kunnen2016}. For most cases, ${\epsilon _{Nu}}$ is less than $1\%$ and the maximum of ${\epsilon _{Nu}}$ is about $1.73\%$, as shown in table \ref{tab1}. 
\subsection{Explored parameter space}
In the present study, we aim at studying the geometric effects on ACRBC systematically. The simulations covered a radius ratio $\eta$ range [0.3, 0.9] and a Ra range [${10^6}$, ${10^8}$]. Pr was fixed at 4.3 corresponding to the working fluids of water at ${40}^{\circ}\mathrm{C}$. $\rm R{o^{ - 1}}$ was fixed at 1, that is because when $\rm R{o^{ - 1}} \ll 1$, the rotation effect is too weak; while when $\rm R{o^{ - 1}}$ $\gg 1$, the turbulence is suppressed, and the flow becomes quasi-2D state. In order to study the joint effect of rotation and buoyancy driving, all cases adopt a moderate $\rm R{o^{ - 1}} = 1$. And besides, all the results in this paper set the angular velocity unit vector $\boldsymbol{\widehat \omega}$ = +1, that is, the system rotates counterclockwise.
\section{Results and discussion}
\begin{figure*}
	\centerline{\includegraphics[width=0.99\linewidth]{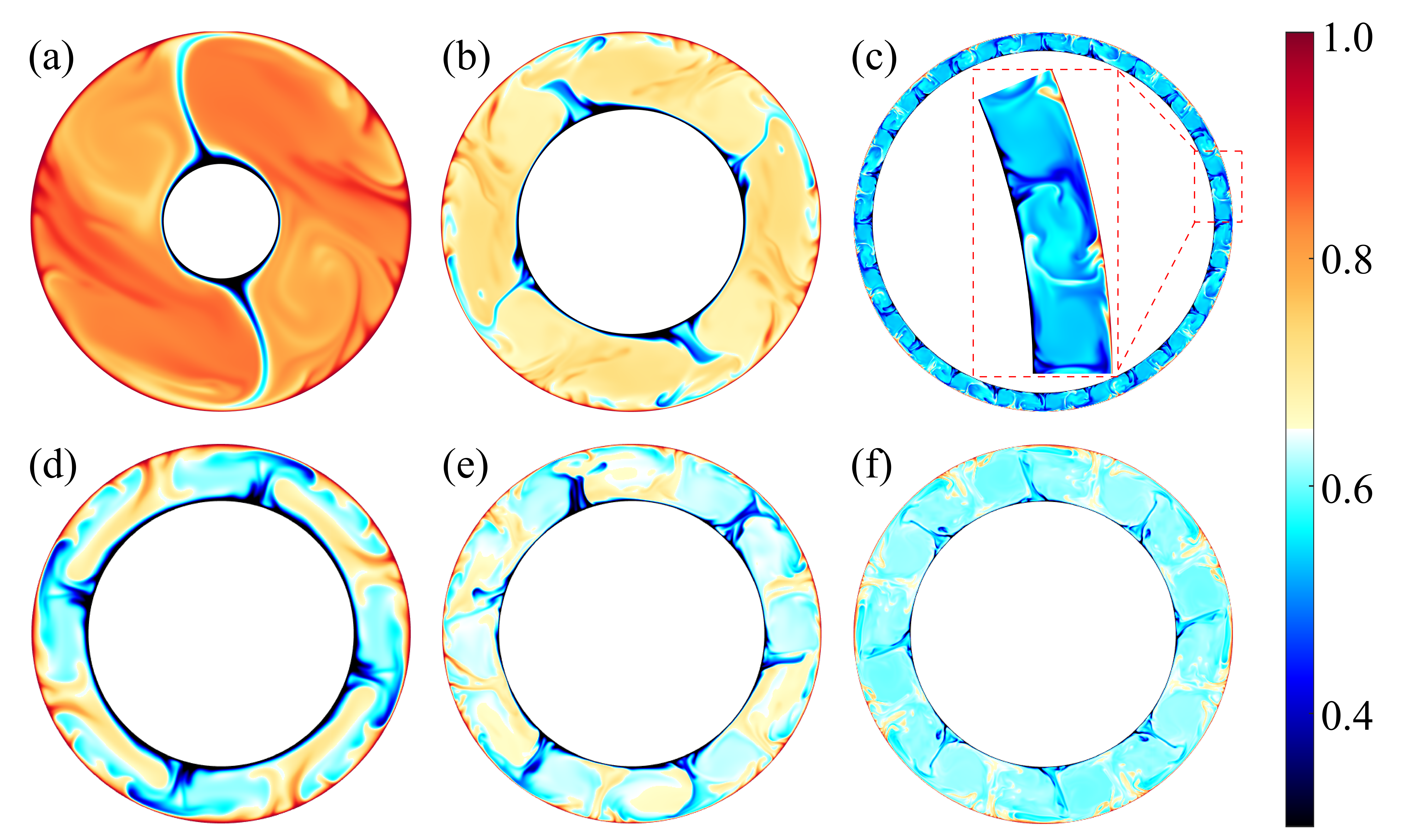}}
	\caption{
		(a-c) Instantaneous temperature fields on a $\phi r$-plane for $\eta$ = 0.3, 0.6, 0.9 at $\rm Ra = {10^7}$, $\rm R{o^{ - 1}} = 1$ and $\Pr  = 4.3$. (d-f) Instantaneous temperature fields for $\rm Ra = {10^6}$, ${10^7}$, ${10^8}$ at $\eta$ = 0.7, $\rm R{o^{ - 1}} = 1$ and $\Pr  = 4.3$. All figures share the same colorbar. 
	}
	\label{Fig3}
\end{figure*}
\subsection{Zonal flow}
As mentioned in \cite{JiangSciAdv}, the convection rolls in ACRBC revolve in prograde direction around the axis with a faster rotation rate than the background rotation of the experimental system, which is the so-called zonal flow. 
The mechanisms of zonal flow and the responses of the dynamics of zonal flow to $\eta$ and Ra are discussed in $\S$ 3.1.1. Next we present some statistical analyses of zonal flow, i.e. the frequency of the net rotation of the convection rolls and the strength of zonal flow $\left\langle {{v_\phi }} \right\rangle_{V,t}$ in  $\S$ 3.1.2, where ${\left\langle {...} \right\rangle _{V,t}}$ denotes the average over time and over whole volume.
\subsubsection{Dynamics of zonal flow}
Akin to the classical RBC, the direction of the temperature gradient is parallel to the centrifugal buoyancy in ACRBC. As the temperature difference or the rotation rate increases, the flow will gradually become unstable and convection appears when Ra is higher than critical $\rm R{a_c}$ \citep{pitz_marxen_chew_2017} of convection onset. When $\rm Ra \gtrsim {10^6}$ \citep{Kang2019PRF}, the flow transitions to be turbulent in the bulk and we will focus on the turbulent convection regime.

In ACRBC, under the rapid rotation of the system, Coriolis force emerges and plays an important role. As shown in figure \ref{Fig2}, driven by the Coriolis force, the cold and hot plumes should both have deflected to the right side from the initial direction. The deflected angle of hot plumes ($\angle \alpha $) is roughly equal to the deflected angle of cold plumes ($\angle \beta $). Considering the centrifugal buoyancy is relatively large near the outer cylinder, the deflected angle is slightly large for the hot plumes.  However, because of the different curvatures of the inner and outer cylinders (see figure \ref{Fig2}(a,c), $\eta$ = 0.4), the hot plumes deflect and then impact close to the region $\rm {A_1}$ where the cold plumes are ejected, whereas the distance between the impacting region $\rm {B_1}$ of the cold plumes and the emitting region of hot plumes is relatively large. Consequently, the hot plumes win and push the overall flow to move counterclockwise. Some cold plumes even deflect to the left side due to the impact of hot plumes. For comparison, the difference between the curvatures of the hot and cold walls is small at $\eta  = 0.9$ (see figure \ref{Fig2}(b,d)) and both the impacting regions $\rm {A_2}$ and $\rm {B_2}$  are far away from the ejecting positions of cold and hot plumes, respectively. Thus, zonal flow becomes weaker with $\eta$ increasing.     Figures \ref{Fig3}(a-c) show the instantaneous temperature fields for different radius ratios $\eta$ at $\rm Ra = {10^7}$. When $\eta  = 0.3$ (figure \ref{Fig3}(a), also see Supplementary Movies 1), the deflected distance of hot plumes are remarkably larger than the cold plumes.  When $\eta  = 0.6$ (figure \ref{Fig3}(b), also see Supplementary Movies 2),  the asymmetric movements of cold and hot plumes still could not be ignored. Hot plumes can impact on the root region of the cold plumes, on the contrary the impacting region of cold plumes cannot affect the hot plumes directly. When $\eta  = 0.9$ (figure \ref{Fig3}(c), also see Supplementary Movies 3), the movements of cold and hot plumes are almost symmetric, and there are several pairs of large scale circulation (LSC) rolls without distinct azimuthal movement, which is similar to the classical RBC.

Figures \ref{Fig3}(d-f) demonstrate the influences of Ra on flow structures for $\eta$ = 0.7. As we know, for classical RBC the aspect ratio of convection rolls approximately equals to one without the confinement effects of sidewalls, which is almost consistent with the case of $\rm Ra = {10^8}$ in ACRBC (see figure \ref{Fig3}(f)). However, we find that the wavenumber decreases with the decrease of Ra. This is because for small Ra case, the centrifugal force is relatively small and the flow is under the control of Coriolis force. Hence, the thermal boundary layers (BLs) develop along the wall and are hard to detach to form a convection roll. On the contrary, the strong centrifugal buoyancy in high Ra case will prompt the detachment of thermal plumes from boundary layers to form more convection rolls. And for the same reason, the deflection of the plume in figure \ref{Fig3}(f) is significantly weaker compared to figure  \ref{Fig3}(d).
\subsubsection{Quantitative analysis of zonal flow}
\begin{figure}
	\centerline{\includegraphics[width=0.99\linewidth]{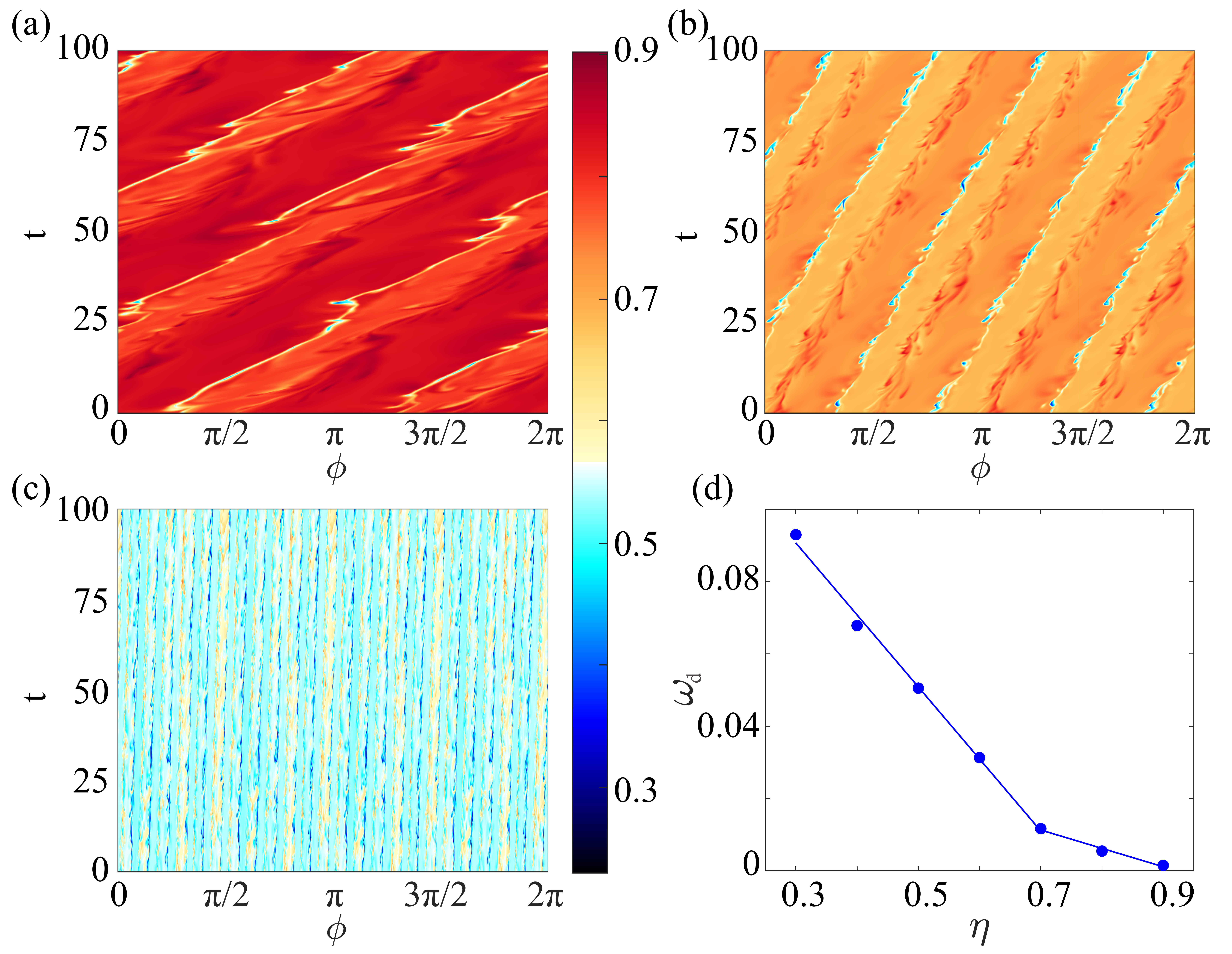}}
	\caption{
		Azimuth-time temperature contours at the mid-radius and mid-axial position for (a) $\eta$ = 0.3, (b) $\eta$ = 0.6 and (c) $\eta$ = 0.9 at $\rm Ra = {10^7}$. All temperature fields share the same colorbar. (d) Drift frequency ${\omega _d}$ of the convection rolls versus $\eta$ at $\rm Ra = {10^7}$, where ${\omega _d}$ is the drift angle (radian) per dimensionless time t. The solid lines  are the best linear fits of ${\omega _d}(\eta)$ for $0.3 \le \eta  \le 0.7$ and $0.7 \le \eta  \le 0.9$, respectively. 
	}
	\label{Fig4}
\end{figure}
Figures \ref{Fig4}(a-c) are the azimuth-time temperature contours at the mid-radius and mid-height position for $\eta$ = 0.3, 0.6 and 0.9, respectively, which show that the convection rolls revolve in prograde direction. And besides, the convection rolls drift at a nearly constant rate with high frequency oscillations.  Similar effects have also been reported in \cite{pitz_marxen_chew_2017}. Figure \ref{Fig4}(d) suggests that the drift frequency ${\omega _d}$ decreases with $\eta$ increasing (averaged over 100 free-fall time units), where ${\omega _d}$ is defined as the azimuthal movement (radian) of convection rolls per free-fall time unit in the rotating frame. It is found that ${\omega _d} =  - 0.05\eta  + 0.05$ for $0.7 \le \eta  \le 0.9$, while it gives a notably steeper slope with  ${\omega _d} =  - 0.2\eta  + 0.15$ for $0.3 \le \eta  \le 0.7$. This change of the slope of ${\omega _d}$ versus $\eta$ indicates the influence of curvature effects on the drift frequency of convection rolls is more significant at small $\eta$.         
\begin{figure}
	\centerline{\includegraphics[width=0.99\linewidth]{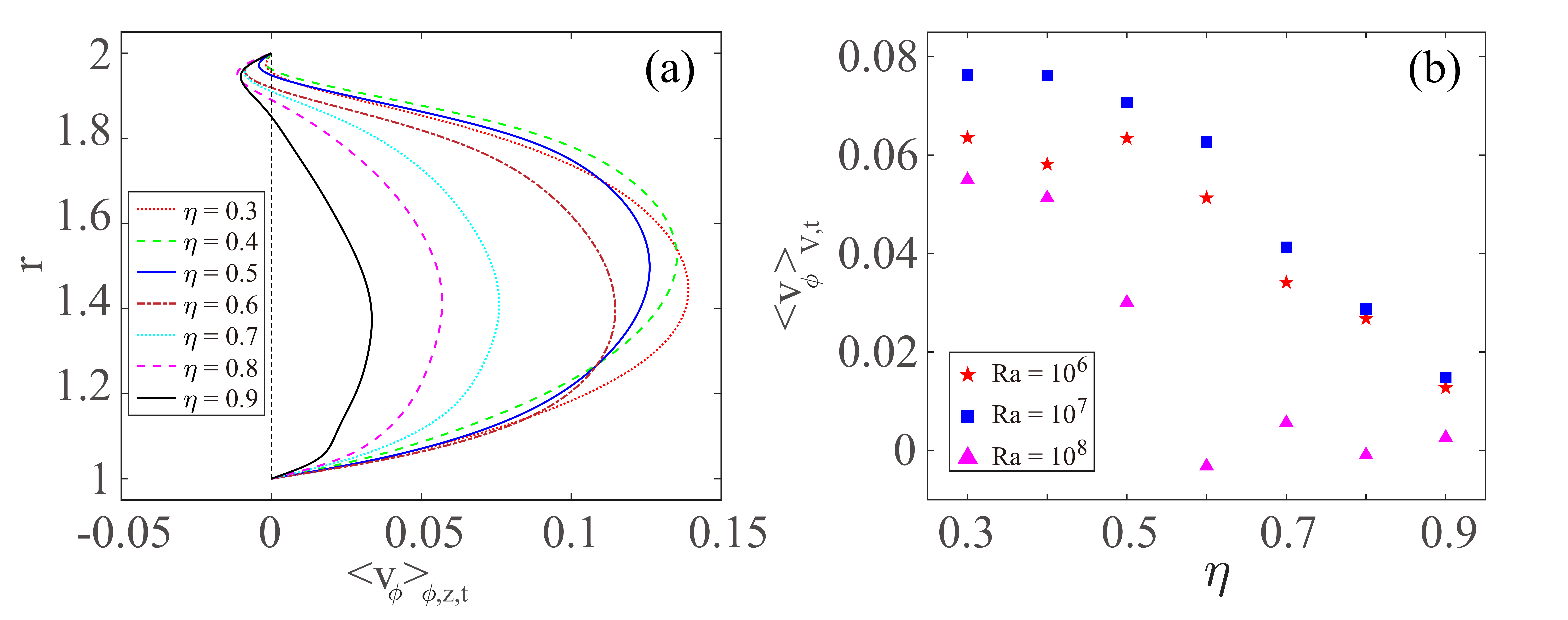}}
	\caption{
		(a) Azimuthal velocity profiles averaged azimuthally, axially and by time along the radial direction for different $\eta$ at $\rm Ra = {10^7}$. (b) Mean azimuthal velocity ${\left\langle {{v_\phi }} \right\rangle _{V,t}}$ as a function of $\eta$ for different Ra. 
	}
	\label{Fig5}
\end{figure} 

Since when there is no systematic deflections of plumes, one would expect that ${\left\langle {{v_\phi }} \right\rangle _{\phi ,z,t}}$ vanishes.   ${\left\langle {{v_\phi }} \right\rangle _{\phi ,z,t}}$ can reflect the asymmetric extent of the motion of cold and hot plumes. As shown in figure \ref{Fig5}(a), with the decrease of $\eta$, the positive ${\left\langle {{v_\phi }} \right\rangle _{\phi ,z,t}}$ (reflects the deflection of hot plumes) is larger and the negative part (reflects the deflection of cold plumes) is smaller in absolute values. Figure \ref{Fig5}(b) shows that for each fixed Ra, space- and time-averaged azimuthal velocity all decrease as $\eta$ increases basically. Thus, the drift of LSC rolls is directly linked to the asymmetric motion of cold and hot plumes. In addition, we find that at moderate Ra ($\rm Ra = {10^7}$), zonal flow is the strongest. When $\rm Ra = {10^6}$, the velocity of plumes is slow due to the weak buoyancy; while when $\rm Ra = {10^8}$, the deflection of plumes is weak as a consequence of strong buoyancy. So the zonal flow driven by the plumes for both cases is relatively weak. 
We also note that when $\eta  \lesssim 0.5$ for $\rm Ra = {10^6}$ and ${10^7}$ or $\eta  \lesssim 0.4$ for $\rm Ra = {10^8}$ (figure \ref{Fig5}(b)),  ${\left\langle {{v_\phi }} \right\rangle _{V,t}}$ no longer increases remarkably as $\eta$ decreases, which is probably a consequence of the increase of the gap width of the convection cell. Long distance to reach the opposite surfaces results in the reduction of the strength of plumes and ${\left\langle {{v_\phi }} \right\rangle _{V,t}}$ reaches saturation.
\subsection{Heat transport}
\begin{figure}
	\centerline{\includegraphics[width=0.99\linewidth]{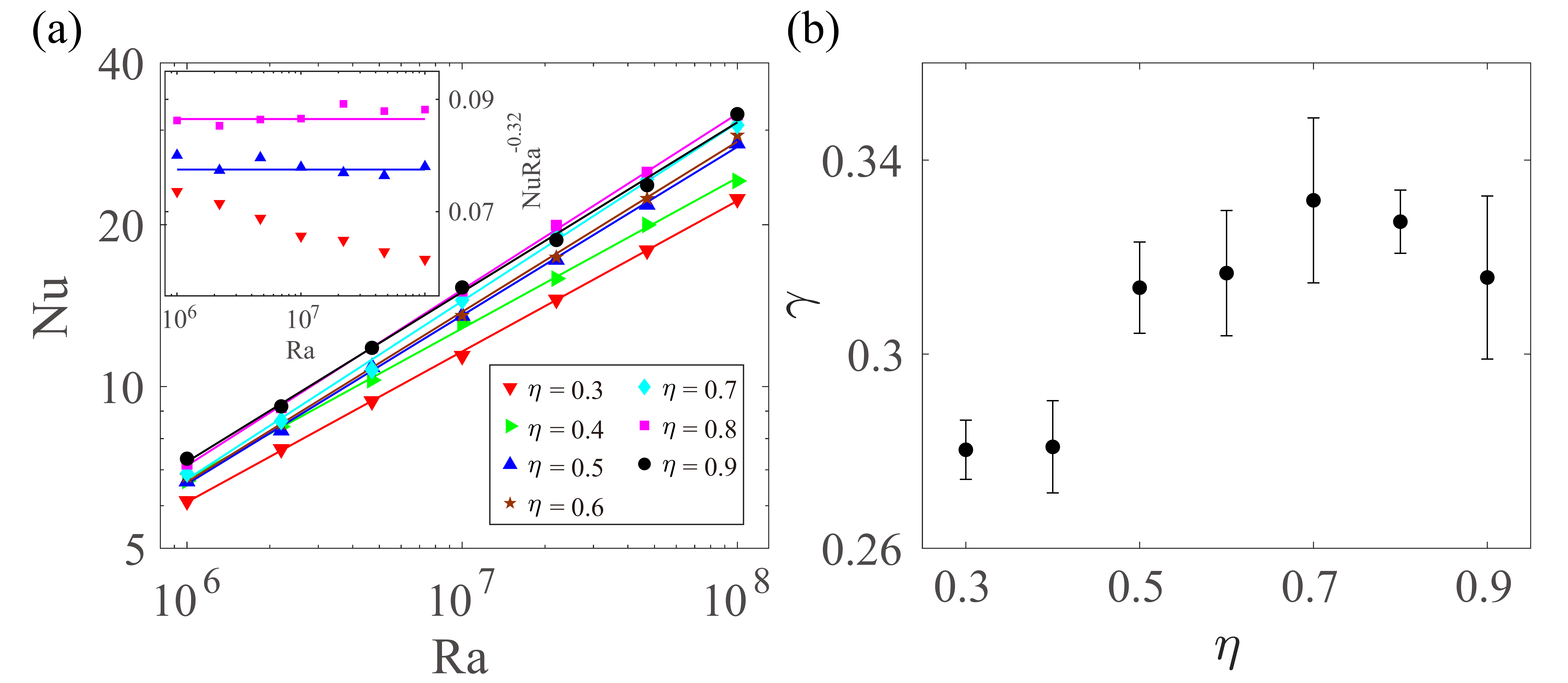}}
	\caption{
		(a) Nusselt number Nu as a function of Rayleigh number Ra for different $\eta$. The inset shows $\rm NuR{a^{ - 0.32}} \sim Ra$ for $\eta$ = 0.3, 0.5, 0.8. (b) Scaling exponents $\gamma$ obtained by fitting $\rm Nu \sim R{a^\gamma }$ as a function of $\eta$. Errorbar is the range of fitting exponent within $95\%$ confidence interval.	
	} 
	\label{Fig6}
\end{figure}
\begin{figure}
	\centerline{\includegraphics[width=0.99\linewidth]{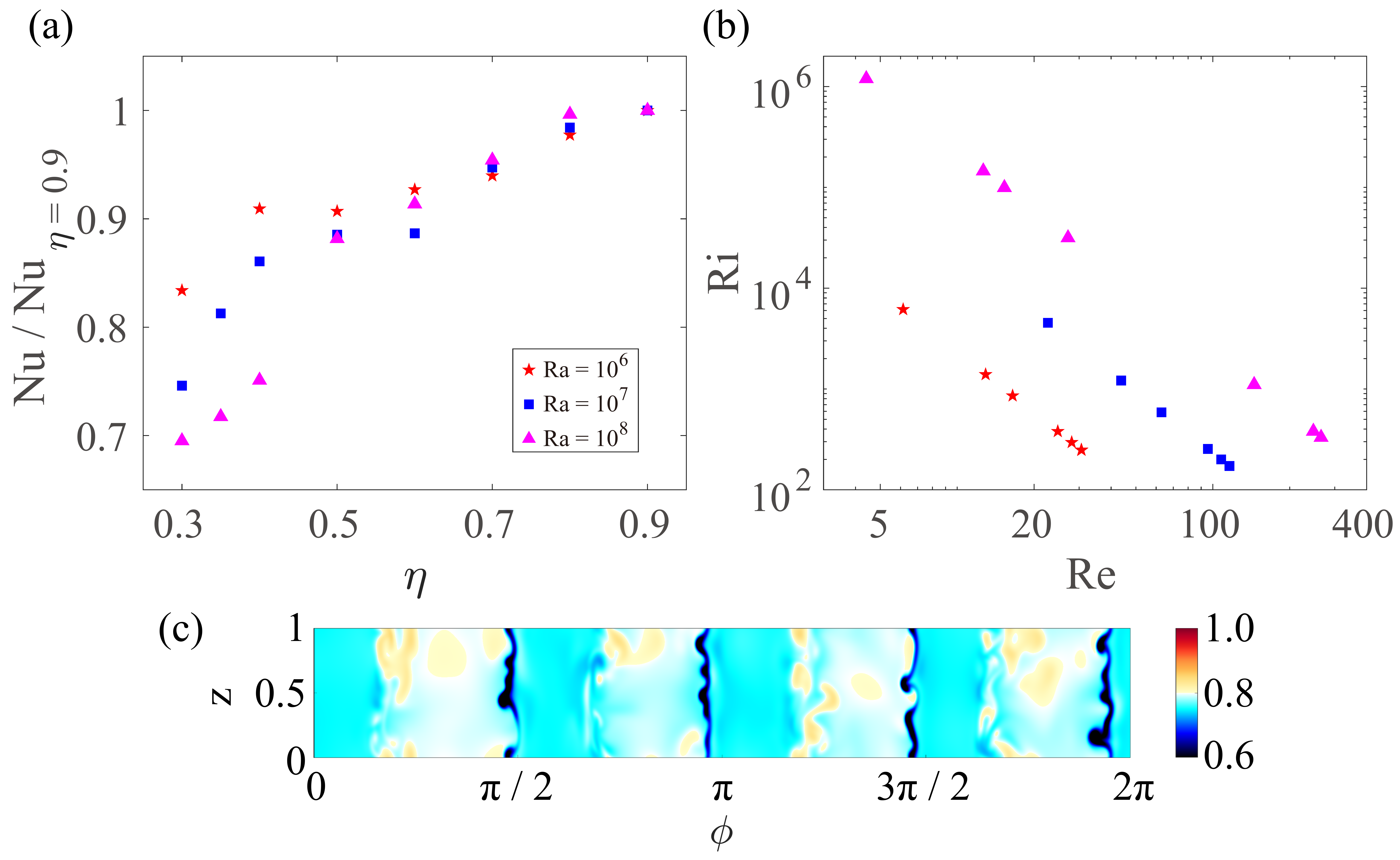}}
	\caption{
		(a) $\rm Nu / N{u_{\eta  = 0.9}}$ as a function of $\eta$ for different Ra. Here $\rm Nu / N{u_{\eta  = 0.9}}$ denotes Nusselt number normalized by the value at $\eta$ = 0.9 for the respective Ra. (b)  Ri versus Re for different Ra. (c) Instantaneous snapshots of temperature field at mid-radius position for Ra = ${10^7}$ and $\eta  = 0.4$.
	}
	\label{Fig7}
\end{figure}

Figure \ref{Fig6}(a) shows Nu as a function of Ra for $0.3 \le \eta  \le 0.9$. When $\eta$ = 0.3 or 0.4, $\rm Nu(Ra)$ can be described with a power law $\rm Nu \sim R{a^\gamma }$ with a scaling exponent $\gamma$ of 0.28 $\pm$ 0.01 (see figure \ref{Fig6}(b)). Whereas for $\eta  \gtrsim 0.5$, it is found that $\gamma$ increases significantly, and then gradually becomes saturated to 0.32 $\pm$ 0.02. Inset of figure \ref{Fig6}(a) shows  $\rm NuR{a^{ - 0.32}}$ as a function of Ra, also suggesting that when $\eta$ = 0.3, $\gamma$ is relatively small.  In Taylor-Couette turbulence, the dimensionless angular velocity flux $\rm N{u_\omega }$ also increases as $\eta$ increases, reaching saturation when $\eta  \gtrsim 0.7$ \citep{Sun2016Annu}. So for flow in rotating cylindrical annulus system, there may exist a unified law that transport efficiency is higher in the system with larger $\eta$.
In figure \ref{Fig7}(a), we plot the normalized Nu (normalized by Nu at $\eta$ = 0.9 for the respective Ra) as a function of $\eta$. It is clearly seen that Nu decreases with the decrease of $\eta$ for each Ra. For the case of $\rm Ra = {10^8}$ and $\eta$ = 0.3, Nu is only about $70\%$ of that at $\eta$ = 0.9.
What is the physical reason for the $\eta$ dependence of heat transport?

As mentioned in $\S$ 3.1, the decrease of $\eta$ results in stronger zonal flow.
One possible reason is that thermal plumes are swept away by the shear of zonal flow. Thus, the heat transport is depressed. We adopt the bulk Richardson number $\rm {Ri = Ra/({\mathop{\rm Re}\nolimits} ^2\Pr )}$ to investigate the competition between the buoyancy and shear effects and analyze the effect of the zonal flow shear by Reynolds number $ \rm {{\mathop{\rm Re}\nolimits}}  $ $= {\left\langle {{u_\phi }} \right\rangle _{V,t}L}/\nu $. As shown in figure \ref{Fig7}(b), it is found that the minimum Ri is 172 corresponding to Ra = ${10^7}$ and $\eta  = 0.4$. From the $\varphi z$ temperature field at the mid-radius position (see figure \ref{Fig7}(c)), it is found that the main flow structure resembles the bulk flow found in RB convection. Note that elongated streaks along the shear flow do not exist for this maximum shear effects case, which suggests that the flow is dominated by the buoyancy. Recently, Blass et al. \cite{Blass2020JFM} have added shearing effects to classical RBC by pulling the top and bottom plates in opposite directions, i.e. the so-called Couette-RB flow. According to the behaviour of the flow structures versus Ri and Re, they considered the flow states are divided into three regimes, including buoyancy dominated regime, transitional regime and shear dominated regime. In ACRBC, Ri is larger than the order of $O$(100) and Re is no more than 300. So the flow is in the buoyancy dominated regime. From the figure 5(b) of \cite{Blass2020JFM}, we can estimate that the reduction of Nu is only about 3\% for $\rm {{{\mathop{\rm Re}\nolimits}}}$ = 300 at Ra = ${10^8}$ by linear interpolation. Thus, the shear effect of zonal flow can inhibit heat transfer, but the effect is very weak. We consider that the decrease of heat transport for small $\eta$ may be due to the geometric confinement effects primarily, which needs to be further studied.
  
\begin{figure}
	\centerline{\includegraphics[width=0.99\linewidth]{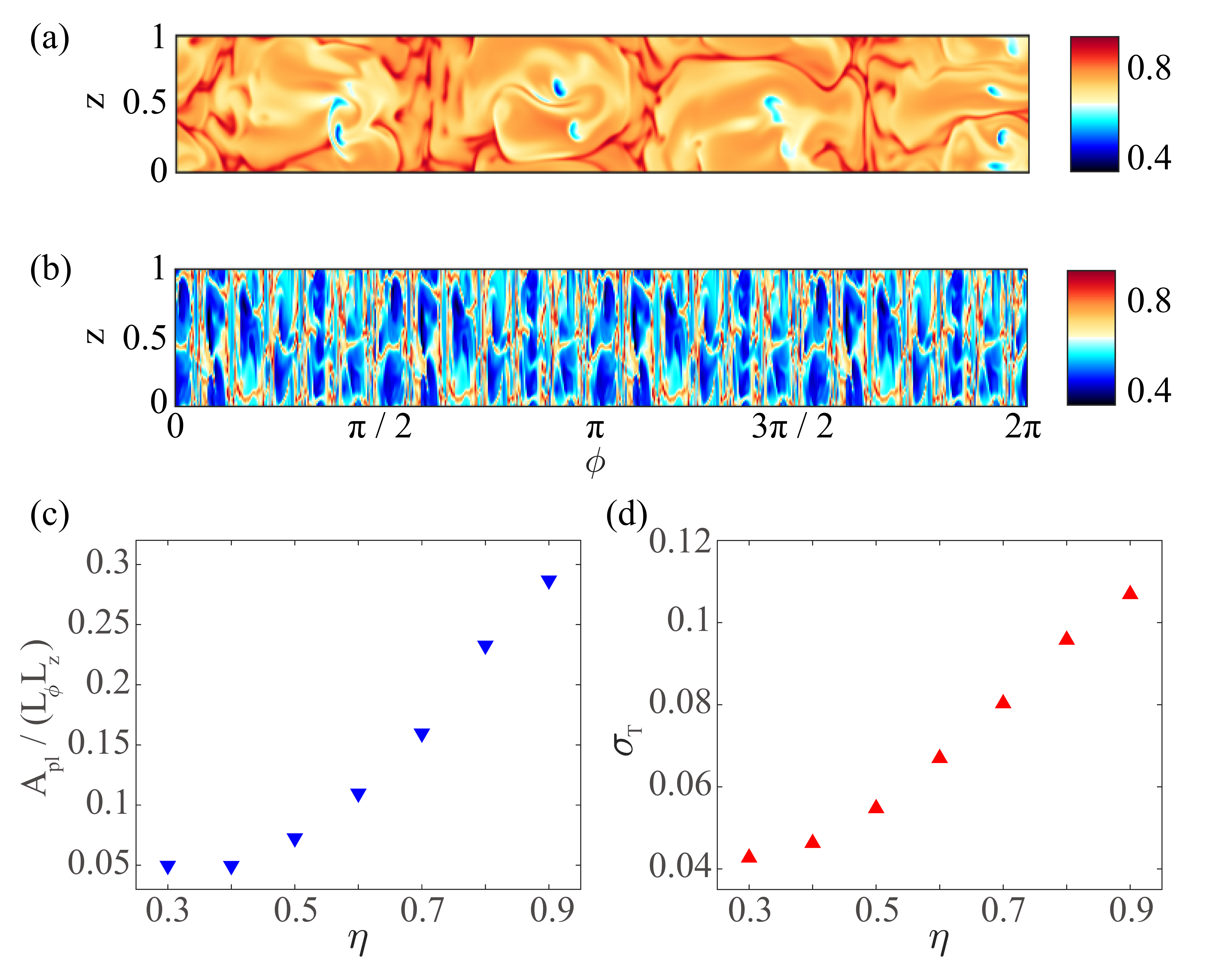}}
	\caption{
		(a) Instantaneous temperature fields in $\phi z$-plane at the edge of the thermal BL (${r_\delta } = {R_o} - {\delta _T}$) for $\eta$ = 0.4 (a), 0.9 (b) at $\rm Ra = {10^7}$, where $\delta_T$ is the thickness of thermal BL obtained using the slope method \citep{gastine_wicht_aurnou_2015}. (a) and (b) share the same colorbar. (c) Ensemble-averaged area of cold plumes normalized by the total area versus $\eta$. (d) Ensemble-averaged standard deviation of temperature at the position of ${r_\delta }$ versus $\eta$.
	}
	\label{Fig8}
\end{figure}

We also investigate the effects of radius ratio on the heat transport from the perspective of plumes. Employing the method introduced in \cite{Huang2013PRL,Poel2015JFM,Chong2017PRL,Jiang2018PRL}, the cold plume coverage at the edge of the thermal BL of the outer cylinder is obtained. Figures \ref{Fig8}(a) and (b) display the temperature field near the outer surface for $\eta$ = 0.4 and 0.9, respectively. It is evident that when $\eta$ = 0.9, the area of cold plumes is larger. Figure \ref{Fig8}(c) shows ensemble-averaged portion of area ${A_{pl}}/({L_\phi }{L_z})$ covered by the cold plumes in the vicinity of the outer surface. As $\eta$ decreases, a smaller portion of cold plumes can travel to the opposing BL. Consequently, temperature fluctuations of the thermal BL of the outer surface are smaller (see figure \ref{Fig8}(d)) and the heat transfer efficiency is declined.

\subsection{Asymmetric mean temperature fields}
In classical RBC,  due to  the Oberbeck-Boussinesq (OB) approximation \citep{Oberbeck1879}, it is generally assumed that the fluid properties are constant, except for the linear variation of the density with temperature in the buoyancy term. However, when the temperature difference $\Delta$ between the hot and cold walls is large enough, the drastic changing of fluid properties should be taken into consideration. In this situation, the top-bottom symmetry of the system is broken and the bulk temperature ${\theta_c}$ deviates from the arithmetic mean temperature of hot and cold walls ${\theta_m}$, which is known as non-Oberbeck-Boussinesq (NOB) effects \citep{Wu1991,NOB2020}.

\begin{figure}
	\centerline{\includegraphics[width=0.99\linewidth]{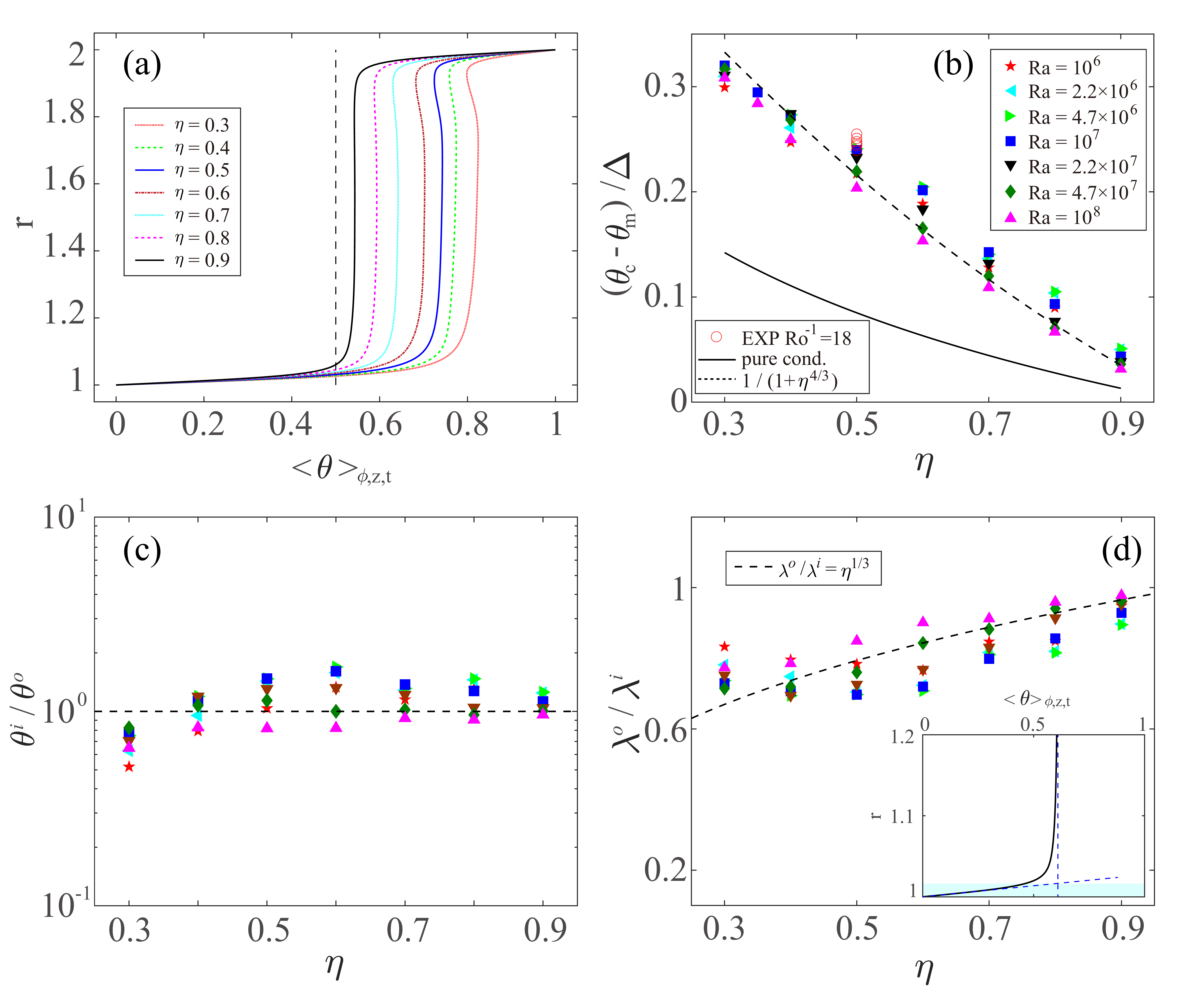}}
	\caption{
		(a) Azimuthally, axially and time-averaged radial temperature profiles for different $\eta$ at $\rm Ra = {10^7}$.
		(b) Relative deviation of the bulk temperature (${\theta_c}$) from arithmetic mean temperature (${\theta_m}$) of the two cylinders $({\theta_c} - {\theta_m})/\Delta$ versus $\eta$ for different Ra from DNS (the solid symbols), experiments (the open symbols), and the theoretical prediction given in (\ref{G}) (dashed line). The experiments are performed at $\eta  = 0.5$, $\rm Ra = 6 \times {10^{10}}$, $\rm Pr  = 4.3$ and $\rm R{o^{ - 1}} = 18$.  The solid line denotes $({\theta_c} - {\theta_m})/\Delta$ in pure thermal conduction state for comparison. (c) Ratio of boundary layer temperature scales ${\theta ^i}/{\theta ^o}$ versus $\eta$ for different Ra. The horizontal dashed line corresponds to the hypothetical identity ${\theta ^i} = {\theta ^o}$ (see (\ref{F})). (d) Ratio of thermal boundary layer thicknesses. The dashed line corresponds to the theoretical prediction given in (\ref{I}). The inset of (d) shows the inner thermal boundary layers  ${\lambda ^i}$ highlighted by the blue shaded area. ${\lambda ^{i,o}}$ are defined as the distances where the tangents of the temperature profiles at the plate cross ${\left\langle \theta  \right\rangle _{\phi ,z,t}} = {\theta _c}$.
	}
	\label{Fig9} 
\end{figure}

Although we implement Oberbeck-Boussinesq approximation in the numerical simulations of ACRBC, remarkable deviation of ${\theta_c}$ from ${\theta_m}$ is observed, which is similar to the NOB effects. Figure \ref{Fig9}(a) shows the radial temperature profiles averaged azimuthally, axially and over time. It is evident that ${\theta_c}$ is much larger than ${\theta_m}$. With $\eta$ increasing, the deviation $({\theta_c} - {\theta_m})/\Delta$ decreases for each Ra (see figure \ref{Fig9}(b)), which illustrates that the mean temperature profile gradually behaves similarly to the classical RBC. We also perform some experiments in ACRBC system.  Details about the experimental setup are introduced in \cite{JiangSciAdv}.    A small thermistor (Measurement Specialties, GAG22K7MCD419, with a response time of 30 ms in liquids) is inserted into the bulk flow to measure ${\theta_c}$. Another type of thermistors (Omega, 44131) evenly distributed along the circumferential direction in the cylinders are used to obtain ${\theta_m}$. $({\theta_c} - {\theta_m})/\Delta$ at $\eta  = 0.5$ for experiments are slightly larger than the cases of DNS, which probably results from the relatively large Ra and $\rm R{o^{ - 1}}$ for experiments ($\rm Ra = 6 \times {10^{10}}$, $\rm Pr  = 4.3$ and $\rm R{o^{ - 1}} = 18$).

Next, we provide a theoretical explanation for the increase of the deviation $({\theta_c} - {\theta_m})/\Delta$ with $\eta$ decreasing.
Since the area of the outer cylinder is larger than that of the inner cylinder, according to Fourier's law, the temperature gradient near the outer wall is smaller. Therefore, ${\theta_c}$ (the temperature at the mid-radius position) is slightly higher than ${\theta_m}$ for pure thermal conduction state in ACRBC, indicated by the solid line in figure \ref{Fig9}(b). While curvature effect of ACRBC accounts for a part of  $({\theta_c} - {\theta_m})/\Delta$, the difference from the total deviation is still finite. To analytically access $({\theta_c} - {\theta_m})/\Delta$  observed in DNS and experiments, we first assume that heat is purely transported by conduction in the thin thermal boundary layers. According to the definition of Nu, heat flux conservation through cylindrical surfaces then yields
\begin{small}
	\begin{align}
	\eta \frac{{\Delta {\theta ^i}}}{{{\lambda ^i}}} &= \frac{{\Delta {\theta ^o}}}{{{\lambda ^o}}} \ ,
	\label{D}
	\end{align}
\end{small}
\noindent where the thermal boundary layer near the inner (outer) wall is assumed to be a conduction profile with a temperature difference $\Delta {\theta ^i}$ ($\Delta {\theta ^o}$) over a thickness ${{\lambda ^i}}$ (${{\lambda ^o}}$). Note that different curvatures of the two cylinders are considered in the definition of Nu.  As shown in figure \ref{Fig9}(a), the bulk fluid is isothermal basically (for small $\eta$, the slight overshoot of temperature near the outer surface may result from the effect of Coriolis effects). Thus, we can further assume that the temperature drops occur only in the boundary layers:
\begin{small}
	\begin{align}
	\Delta {\theta ^i} + \Delta {\theta ^o} &= 1 \ .
	\label{E}
	\end{align}
\end{small}
Equations (\ref{D}) and (\ref{E}) are not sufficient to determine the three unknowns $\Delta {\theta ^i}$, $\Delta {\theta ^o}$ and ${\lambda ^i}/{\lambda ^o}$. So an additional physical assumption is required.

Wu \& Libchaber \cite{Wu1991}, Zhang et al. \cite{Zhang1997}, Gastine et al. \cite{gastine_wicht_aurnou_2015} proposed that the thermal boundary layers adjust their length scales so that the mean hot and cold temperature fluctuations are equal in the bulk region. Temperature fluctuations in the bulk region are caused by plumes detaching from the boundary layers. So this implies the equality of the temperature scales for the two boundary layers (${\theta ^{i,o}} \equiv (\nu \kappa )/(\alpha {g_{i,o}}{({\lambda ^{i,o}})^3})$). This third assumption yields
\begin{small}
	\begin{align}
	{\theta ^i} &= {\theta ^o} \to \frac{{\nu \kappa }}{{\alpha {g_i}{{({\lambda ^i})}^3}}} = \frac{{\nu \kappa }}{{\alpha {g_o}{{({\lambda ^o})}^3}}} \ ,
	\label{F}
	\end{align}
\end{small}
\noindent where the centrifugal acceleration $g = {\omega ^2}r$, so ${\chi _g} = g({R_i})/g({R_o}) = {\eta}$ reflects radially-dependent gravity effects. Figure \ref{Fig9}(c) shows ${\theta ^i}/{\theta ^o}$ for different $\eta$ and Ra. Compared to turbulent RB convection in spherical shells reported by Gastine et al.  \cite{gastine_wicht_aurnou_2015} (see their figure 5(a)), the identity of the boundary layer temperature scale is better fulfilled in ACRBC. The temperature drops at both boundaries and the ratio of the thermal BL thicknesses can then be derived using (\ref{D})-(\ref{F}) 
\begin{small}
	\begin{align}
	\Delta {\theta ^i} \simeq {\theta _c} &= \frac{1}{{1 + {\chi _g}^{1/3}\eta }} = \frac{1}{{1 + {\eta ^{4/3}}}} \ ,
	\label{G}
	\end{align}
\end{small}
\begin{small}
	\begin{align}
	\Delta {\theta ^o} &= \frac{1}{{1 + {\eta ^{ - 4/3}}}} \ ,
	\label{H}
	\end{align}
\end{small}
\begin{small}
	\begin{align}
	{\lambda ^o}/{\lambda ^i} &= {\eta ^{1/3}} \ ,
	\label{I}
	\end{align}
\end{small}
\noindent Figure \ref{Fig9}(b) shows that the predicted value (\ref{G}) and the actual values for mean bulk temperature are in good agreement. We adopt the slope method to define the thickness of thermal boundary layer $\lambda$ as the distance where the tangent of the temperature profiles at the plate intersects ${\left\langle \theta  \right\rangle _{\phi ,z,t}} = {\theta _c}$ (see the inset of figure \ref{Fig9}(d)). Figure \ref{Fig9}(d) shows that the theory (\ref{I}) can also predict the asymmetry of the thermal boundary layers ${\lambda ^o}/{\lambda ^i}$. In a word, the theory accurately accounts for the bulk temperature and the boundary layer asymmetry for different $\eta$ observed in ACRBC. The asymmetry is likely caused by the curvature effects (\ref{D}) and radially-dependent centrifugal force (\ref{F}). We note that, although  asymmetric mean temperature fields exist in ACRBC, their influence on global heat transport has been shown to be negligible \citep{JiangSciAdv}.
\section{Conclusion}
We present an analysis of the zonal flow, heat transport and asymmetric mean temperature field in Annular Centrifugal Rayleigh-B\'{e}nard Convection, by means of high-resolution 3D DNS, with radius ratio $\eta$ varying from 0.3 to 0.9, Ra varying from ${10^6}$ to ${10^8}$, and Pr, $\rm R{o^{ - 1}}$ fixed at 4.3 and 1, respectively.  Major findings are summarized as follows:

Firstly, convection rolls move counterclockwise around the axis with a faster rotation rate than the background rotation of ACRBC system.  Coriolis force results in the similar deflected angle of hot and cold plumes.   Because of the different curvatures of the inner and outer cylinders, the hot plumes can directly affect the ejecting position of the cold plumes, while the impact of the cold plumes does not directly affect the hot plumes. The asymmetric motions of cold and hot plumes push the overall flow to move in the same direction of the background rotation.  This mechanism of zonal flow is verified by the variation of the flow behaviour with radius ratio.
We observe the drift frequency decreases with $\eta$ increasing. It is also found that ${\left\langle {{v_\phi }} \right\rangle _{V,t}}$ decreases as $\eta$ increases for each Ra. In addition, zonal flow is the strongest at moderate Ra ($\rm Ra = {10^7}$), which suggests zonal flow is related to the balance of Coriolis force and buoyancy.  

Secondly, the scaling exponent of global heat transport $\gamma$ is 0.28$\pm$0.01 for $\eta$ = 0.3, 0.4, and transitions to 0.32$\pm$0.02 for $\eta  \gtrsim 0.5$. The dependence of transport efficiency on $\eta$ is consistent with Taylor-Couette flow. Furthermore, $\rm Nu/N{u_{\eta  = 0.9}}$ decreases with the decrease of $\eta$ for each Ra. By analyzing the flow field and the ratio of buoyancy and shear effects, it is found that the flow is in the buoyancy dominated regime and the influence of shear effects of zonal flow on heat transfer is not strong. Additionally, we observed that the cold plume coverage at the edge of the thermal BL of the outer cylinder becomes smaller as $\eta$ decreases.

Thirdly, the bulk temperature ${\theta_c}$ is found to be much higher than ${\theta_m}$ (0.5) by DNS and experiments. The relative deviation $({\theta_c} - {\theta_m})/\Delta$ increases with the decrease of $\eta$ for each Ra. The results of numerical simulations and experiments are consistent, even though the values of Ra and $\rm R{o^{ - 1}}$ are distinct. By assuming pure conduction in the thermal BLs, isothermal bulk fluid, and the equality of the temperature scales for the two boundary layers, we analytically obtain the mean bulk temperature and the ratio of the thickness of the BLs, which are in good agreement with the actual values. The asymmetric mean temperature fields may result from curvature effects and the radially-dependent centrifugal force involved in the hypotheses of the theory.

In the laboratory experiment, centrifugal buoyancy can be larger than Earth's gravity by the rapid rotation of ACRBC.
Hypergravitational thermal convection is a new method to achieve high Ra \citep{JiangSciAdv}. Systematic study on the  dependence of turbulent flow structures and heat transport on radius ratio is urgently needed to understand the flow dynamics in ACRBC. The physical mechanism of zonal flow may improve understanding of some flow phenomena in astrophysical and geophysical settings. Furthermore, the bulk temperature can be significantly increased in the hypergravitational thermal convection system by reducing $\eta$, which give insights for the flow and temperature control in engineering applications.
\section*{Acknowledgements}
This work was supported by the Natural Science Foundation of China (Grant Nos. 11988102, 91852202, 11861131005) and Tsinghua University Initiative Scientific Research Program (Grant No. 20193080058).
\section*{Appendix}
Table \ref{tab1} provide the simulation parameters which are illustated in the subsection of 'Direct numerical simulations'. Movie1, Movie2 and Movie3 are the temperature fields on a $\phi r$-plane at the mid-axial position for $\eta$  = 0.3, 0.6 and 0.9 at $\rm Ra = {10^7}$, $\rm R{o^{ - 1}} = 1$ and $\Pr  = 4.3$,  respectively.
\clearpage

\begin{table*}
	\begin{center}
			\renewcommand\tabcolsep{5pt} 
			\renewcommand\arraystretch{0.5}   
		\begin{tabular}{cccccccccccc}
            \toprule
			$No.$ & $\eta$     & $\text{Ra}$        & ${\Delta _g}/{\eta _K}$       & ${N_{tBL}}$     & ${N_{vBL}}$        & ${\tau _{avg}}$     & $\rm Nu$     & ${\epsilon _{\rm Nu}}$     & $\Gamma$     & ${\phi _0}$     & ${N_\phi } \times {N_z} \times {N_r}$ \\  
			\colrule
			1 & 0.3   & $1.0\times{10^6}$   & 0.24  & 16    & 17    & 255   & 6.12  & 0.67\% & 1     & 1     & $1024\times128\times128$ \\
			2 & 0.3   & $2.2\times{10^6}$  & 0.31  & 13    & 12    & 152   & 7.68  & 1.09\% & 1     & 1     & $1024\times128\times128$ \\
			3 & 0.3   & $4.7\times{10^6}$   & 0.40  & 11    & 13      & 377   & 9.39  & 0.90\% & 1     & 1     & $1024\times128\times128$\\
			4 & 0.3   & $1.0\times{10^7}$   & 0.51  & 12    & 17   & 247   & 11.41  & 0.79\% & 1     & 1     & $1024\times128\times128$\\
			5 & 0.3   & $2.2\times{10^7}$ & 0.46  & 17    & 22     & 105   & 14.46  & 0.57\% & 0.5   & 1/2   & $960\times96\times192$\\
			6 & 0.3   & $2.2\times{10^7}$ & 0.35    & 24      & 25       & 108   & 14.20  & 1.36\% & 0.25   & 1/2   & $1024\times64\times256$\\
			7 & 0.3   & $4.7\times{10^7}$ & 0.58    &  15     & 17       & 181   & 17.88  & 0.15\% & 0.5     & 1/2   &$960\times96\times192$\\
			8 & 0.3   & $1.0\times{10^8}$  & 0.57  & 18    & 19     & 101   & 22.37  & 1.26\% & 0.25  & 1/2   & $1024\times64\times256$ \\
			\hline
			9 & 0.4   & $1.0\times{10^6}$   & 0.25  & 16    & 13     & 122   & 6.69  & 1.23\% & 1     & 1     & $1408\times128\times128$ \\
			10 & 0.4   & $2.2\times{10^6}$  & 0.32  & 14    & 16     & 282   & 8.42  & 0.70\% & 1     & 1     & $1408\times128\times128$ \\
			11 & 0.4   & $4.7\times{10^6}$   & 0.41  & 13    & 16     & 179   & 10.28  & 0.90\% & 1     & 1     & $1408\times128\times128$ \\
			12 & 0.4   & $1.0\times{10^7}$   & 0.53  & 11    & 13    & 1306  & 13.14  & 0.01\% & 1     & 1     &$1408\times128\times128$ \\
			13 & 0.4   & $2.2\times{10^7}$   & 0.47  & 16    & 19     & 205   & 15.88  & 0.60\% & 0.5   & 1     & $2112\times96\times192$ \\
			14 & 0.4   & $4.7\times{10^7}$ & 0.61  & 14    & 17     & 641   & 19.99  & 0.58\% & 0.5   & 1     &$2112\times96\times192$  \\
			15 & 0.4   & $1.0\times{10^8}$  & 0.59  & 17    & 23     & 136   & 23.95  & 1.51\% & 0.5   & 1/2   & $1280\times128\times256$ \\
			\hline
			16 & 0.5   & $1.0\times{10^6}$  & 0.25  & 18    & 19     & 105   & 6.66  & 1.33\% & 1     & 1     & $1536\times128\times128$ \\
			17 & 0.5   & $2.2\times{10^6}$   & 0.40  & 11    & 13    & 521   & 8.28  & 0.49\% & 1     & 1     & $1212\times102\times102$\\
			18 & 0.5   & $4.7\times{10^6}$   & 0.52  & 9     & 12     & 786   & 10.86  & 0.12\% & 1     & 1     & $1212\times102\times102$ \\
			19 & 0.5   & $1.0\times{10^7}$   & 0.54  & 11    & 13      & 200   & 13.54  & 0.20\% & 1     & 1     & $1536\times128\times128$\\
			20 & 0.5   & $2.2\times{10^7}$   & 0.56  & 12    & 16     & 333   & 17.21  & 0.74\% & 1     & 1     & $1944\times162\times162$ \\
			21 & 0.5   & $4.7\times{10^7}$   & 0.62  & 13    & 17     & 540   & 21.79  & 0.25\% & 0.5   & 1/2   &$1152\times96\times192$  \\
			22 & 0.5   & $1.0\times{10^8}$  & 0.62  & 15    & 20     & 166   & 28.33  & 0.59\% & 0.25  & 1/2   &$1536\times64\times256$  \\
			23 & 0.5   & $1.0\times{10^8}$  & 0.62  & 15    & 20       & 711   & 28.04  & 1.45\% & 1     & 1     &$3072\times256\times256$  \\
			\hline
			24 & 0.6   & $1.0\times{10^6}$   & 0.25  & 17    & 16     & 416   & 6.80  & 0.31\% & 1     & 1/2   & $1024\times128\times128$ \\
			25 & 0.6   & $2.2\times{10^6}$   & 0.33  & 15    & 16     & 381   & 8.52  & 0.05\% & 1     & 1/2   & $1024\times128\times128$ \\
			26 & 0.6   & $4.7\times{10^6}$  & 0.42  & 13    & 15     & 351   & 10.68  & 0.49\% & 1     & 1/2   & $1024\times128\times128$ \\
			27 & 0.6   & $1.0\times{10^7}$   & 0.55  & 11    & 13     & 654   & 13.56  & 0.21\% & 1     & 1/2   & $1024\times128\times128$ \\
			28 & 0.6   & $2.2\times{10^7}$   & 0.49  & 15    & 20     & 341   & 17.40  & 1.05\% & 1     & 1/2   &$1536\times192\times192$  \\
			29 & 0.6   & $4.7\times{10^7}$   & 0.64  & 12    & 18    & 389   & 22.38  & 0.33\% & 1     & 1/2   & $1536\times192\times192$ \\
			30 & 0.6   & $1.0\times{10^8}$  & 0.63  & 15    & 21     & 305   & 29.35  & 0.14\% & 0.25  & 1/4   & $1024\times64\times256$ \\
			31 & 0.6   & $1.0\times{10^8}$  & 0.64  & 14    &  19    & 161   & 30.95  & 0.90\% & 0.25  & 1/2   & $2048\times64\times256$ \\
			\hline
			32 & 0.7   & $1.0\times{10^6}$   & 0.25  & 17    & 20     & 714   & 6.90  & 0.06\% & 1     & 1/4   & $768\times128\times128$ \\
			33 & 0.7   & $2.2\times{10^6}$   & 0.33  & 15    & 17     & 528   & 8.63  & 0.47\% & 1     & 1/4   & $768\times128\times128$ \\
			34 & 0.7   & $4.7\times{10^6}$  & 0.42  & 13    & 16     & 711   & 10.74  & 0.07\% & 1     & 1/4   & $768\times128\times128$ \\
			35 & 0.7   & $1.0\times{10^7}$  & 0.56  & 10    & 14      & 664   & 14.48  & 0.36\% & 1     & 1     &$2560\times128\times128$  \\
			36 & 0.7   & $2.2\times{10^7}$   & 0.51  & 14    & 19     & 349   & 19.22  & 0.04\% & 1     & 1/4   &$1152\times192\times192$  \\
			37 & 0.7   & $4.7\times{10^7}$   & 0.65  & 12    & 16     & 439   & 24.29  & 1.19\% & 1     & 1/4   & $1152\times192\times192$ \\
			38 & 0.7   & $1.0\times{10^8}$  & 0.64  & 15    & 22     & 145   & 30.65  & 0.62\% & 1     & 1/4   &$1280\times256\times256$  \\
			\hline
			39 & 0.8   & $1.0\times{10^6}$   & 0.39  & 10    & 12     & 1013  & 7.17  & 0.03\% & 1     & 1     & $2512\times82\times82$ \\
			40 & 0.8   & $2.2\times{10^6}$  & 0.50  & 8     & 10      & 678   & 9.13  & 0.04\% & 1     & 1     & $2512\times82\times82$ \\
			41 & 0.8   & $4.7\times{10^6}$  & 0.54  & 9     & 10     & 228   & 11.79  & 0.65\% & 1     & 1     & $3132\times102\times102$ \\
			42 & 0.8   & $1.0\times{10^7}$  & 0.56  & 10    & 13      & 461   & 15.05  & 0.13\% & 1     & 1     & $3968\times128\times128$\\
			43 & 0.8   & $2.2\times{10^7}$   & 0.49  & 15    & 20     & 228   & 19.95  & 0.86\% & 0.5   & 1/4   & $1624\times102\times204$ \\
			44 & 0.8   & $4.7\times{10^7}$   & 0.62  & 12    & 18      & 249   & 25.06  & 0.46\% & 0.5   & 1/4   & $1624\times102\times204$ \\
			45 & 0.8   & $1.0\times{10^8}$  & 0.65  & 14    & 21     & 381   & 32.01  & 0.51\% & 0.25  & 1/4   &$2048\times64\times256$  \\
			\hline
			46 & 0.9   & $1.0\times{10^6}$   & 0.26  & 17    & 19    & 199   & 7.31  & 1.73\% & 1     & 1/8   & $1024\times128\times128$ \\
			47 & 0.9   & $2.2\times{10^6}$  & 0.34  & 15    & 17      & 223   & 9.19  & 1.50\% & 1     & 1/8   &$1024\times128\times128$  \\
			48 & 0.9   & $4.7\times{10^6}$   & 0.44  & 12    & 17      & 233   & 11.80  & 1.36\% & 1     & 1/8   & $1024\times128\times128$ \\
			49 & 0.9   & $1.0\times{10^7}$  & 0.56  & 10    & 13     & 178   & 15.29  & 0.94\% & 1     & 1/4   & $1984\times128\times128$ \\
			50 & 0.9   & $2.2\times{10^7}$  & 0.50  & 14    & 19      & 363   & 18.74  & 0.16\% & 1     & 1/8   & $1536\times192\times192$ \\
			51 & 0.9   & $4.7\times{10^7}$   & 0.65  & 12    & 19    & 499   & 23.70  & 0.23\% & 1     & 1/8   & $1536\times192\times192$ \\
			52 & 0.9   & $1.0\times{10^8}$  & 0.65  & 14    & 21    & 85    & 32.13  & 0.11\% & 0.25  & 1/4   & $4096\times64\times256$ \\
			\botrule
		\end{tabular}
		
		\caption{Simulation parameters. The columns from left to right indicate the followings: radius ratio $\eta$, Rayleigh number Ra, the maximum grid spacing ${\Delta _g}$ compared with the Kolmogorov scale estimated by the global criterion ${\eta _K}{\rm{ = }}\frac{{L{{\Pr }^{1/2}}}}{{[\rm{Ra(Nu - 1)}]}^{1/4}} \cdot {[\frac{{(1 + \eta )\ln (\eta )}}{{2(\eta  - 1)}}]^{1/4}}$ \citep{JiangSciAdv}, the number of grid points within the thermal BL ${N_{tBL}}$ and  viscous BL ${N_{vBL}}$, the averaging time period ${\tau _{avg}}$, Nusselt number Nu, the difference of Nu between inner and outer walls ${\epsilon _{Nu}} = \left| \rm {{N{u_{in}} - N{u_{out}}}} \right|/\rm {Nu}$, aspect ratio $\Gamma$, reduced  azimuthal  domain ${\phi _0}$, the resolution in azimuthal, axial, and radial directions ${N_\phi } \times {N_z} \times {N_r}$. }
		\label{tab1}
	\end{center}
\end{table*}

\clearpage

\bigskip


\begin{thebibliography}{47}

\expandafter\ifx\csname natexlab\endcsname\relax\def\natexlab#1{#1}\fi
\def\au#1{#1} \def\ed#1{#1} \def\yr#1{#1}\def\at#1{#1}\def\jt#1{\textit{#1}}
\def\bt#1{#1}\def\bvol#1{\textbf{#1}} \def\vol#1{#1} \def\pg#1{#1}
\def\publ#1{#1}\def\arxiv#1{#1}\def\org#1{#1}\def\st#1{\textit{#1}}

\bibitem[Mckenzie {\em et~al.\/}(1974)Mckenzie, Roberts \& Weiss]{mantle}
{\sc \au{Mckenzie, D.P.}, \au{Roberts, J.M.} \& \au{Weiss, N.O.}} \yr{1974}
\at{{Convection in the earth's mantle: towards a numerical simulation}}.
\jt{J. Fluid Mech.}  \bvol{62}~(3),  \pg{465–538}.

\bibitem[Cardin \& Olson(1994)]{outer_core}
{\sc \au{Cardin, P.} \& \au{Olson, P.}} \yr{1994}  \at{Chaotic thermal
	convection in a rapidly rotating spherical shell: consequences for flow in
	the outer core}.  \jt{Phys. Earth Planet. Inter.}  \bvol{82}~(3),  \pg{235 --
	259}.

\bibitem[Wyngaard(1992)]{atmosphere1}
{\sc \au{Wyngaard, J.C.}} \yr{1992}  \at{{Atmospheric Turbulence}}.  \jt{Annu.
	Rev. Fluid Mech.}  \bvol{24}~(1),  \pg{205--234}.

\bibitem[Hartmann {\em et~al.\/}(2001)Hartmann, Moy \& Fu]{atmosphere2}
{\sc \au{Hartmann, D.L.}, \au{Moy, L.A.} \& \au{Fu, Q.}} \yr{2001}
\at{{Tropical Convection and the Energy Balance at the Top of the
		Atmosphere}}.  \jt{J. Clim.}  \bvol{14}~(24),  \pg{4495}.

\bibitem[Cheng {\em et~al.\/}(2019)Cheng, Abraham, Hausfather \&
Trenberth]{ocean}
{\sc \au{Cheng, L.}, \au{Abraham, J.}, \au{Hausfather, Z.} \& \au{Trenberth,
		K.E.}} \yr{2019}  \at{{How fast are the oceans warming?}}  \jt{Science}
\bvol{363},  \pg{128--129}.

\bibitem[Michael~Owen \& Long(2015)]{engineering}
{\sc \au{Michael~Owen, J.} \& \au{Long, C.A.}} \yr{2015}  \at{{Review of
		Buoyancy-Induced Flow in Rotating Cavities}}.  \jt{J. Turbomach.}
\bvol{137},  \pg{11}.

\bibitem[Hide \& Mason(1975)]{Hide1975}
{\sc \au{Hide, R.} \& \au{Mason, P.J.}} \yr{1975}  \at{{Sloping convection in a
		rotating fluid}}.  \jt{Adv. Phys.}  \bvol{24}~(1),  \pg{47--100}.

\bibitem[Bohn {\em et~al.\/}(1995)Bohn, Deuker, Emunds \& Gorzelitz]{Bohn1995}
{\sc \au{Bohn, D.}, \au{Deuker, E.}, \au{Emunds, R.} \& \au{Gorzelitz, V.}}
\yr{1995}  \at{{Experimental and Theoretical Investigations of Heat Transfer
		in Closed Gas-Filled Rotating Annuli}}.  \jt{J. Turbomach.}  \bvol{117}~(1),
\pg{175--183}.

\bibitem[Ahlers {\em et~al.\/}(2009)Ahlers, Grossmann \& Lohse]{Review1}
{\sc \au{Ahlers, G.}, \au{Grossmann, S.} \& \au{Lohse, D.}} \yr{2009}
\at{{Heat transfer and large scale dynamics in turbulent
		{{Rayleigh-B\'enard}} convection}}.  \jt{Rev. Mod. Phys.}  \bvol{81},
\pg{503--537}.

\bibitem[Lohse \& Xia(2010)]{Review2}
{\sc \au{Lohse, D.} \& \au{Xia, K.-Q.}} \yr{2010}  \at{{Small-Scale Properties
		of Turbulent {{Rayleigh-B\'enard}} Convection}}.  \jt{Annu. Rev. Fluid Mech.}
\bvol{42}~(1),  \pg{335--364}.

\bibitem[Chillà \& Schumacher(2012)]{Review3}
{\sc \au{Chillà, F.} \& \au{Schumacher, J.}} \yr{2012}  \at{{New perspectives
		in turbulent {{Rayleigh-B\'enard}} convection.}}  \jt{Eur. Phys. J. E}
\bvol{35}~(7),  \pg{1 -- 25}.

\bibitem[Xia(2013)]{Review4}
{\sc \au{Xia, K.-Q.}} \yr{2013}  \at{Current trends and future directions in
	turbulent thermal convection}.  \jt{Theor. and Appl. Mech. Lett.}
\bvol{3}~(5),  \pg{052001}.

\bibitem[Zou {\em et~al.\/}(2019)Zou, Zhou, Chen, Bao, Chen \& She]{Acta1}
{\sc \au{Zou, H.-Y.}, \au{Zhou, W.-F.}, \au{Chen, X.}, \au{Bao, Y.}, \au{Chen,
		J.} \& \au{She, Z.-S.}} \yr{2019}  \at{{Boundary layer structure in turbulent
		{{Rayleigh-B\'enard}} convection in a slim box}}.  \jt{Acta Mech. Sin.}
\bvol{35},  \pg{713--728}.

\bibitem[Yu {\em et~al.\/}(2019)Yu, Liu, Zhou, Gao \& Liu]{Acta2}
{\sc \au{Yu, Y.}, \au{Liu, F.}, \au{Zhou, T.}, \au{Gao, C.} \& \au{Liu, Y.}}
\yr{2019}  \at{{Numerical solutions of 2-D steady compressible natural
		convection using high-order flux reconstruction}}.  \jt{Acta Mech. Sin.}
\bvol{35},  \pg{401--410}.

\bibitem[Chen {\em et~al.\/}(2020)Chen, Wang \& Xi]{chen_wang_xi_2020}
{\sc \au{Chen, X.}, \au{Wang, D.-P.} \& \au{Xi, H.-D.}} \yr{2020}  \at{{Reduced
		flow reversals in turbulent convection in the absence of corner vortices}}.
\jt{J. Fluid Mech.}  \bvol{891},  \pg{R5}.

\bibitem[Kang {\em et~al.\/}(2019)Kang, Meyer, Yoshikawa \&
Mutabazi]{Kang2019PRF}
{\sc \au{Kang, C.}, \au{Meyer, A.}, \au{Yoshikawa, H.N.} \& \au{Mutabazi, I.}}
\yr{2019}  \at{{Numerical study of thermal convection induced by centrifugal
		buoyancy in a rotating cylindrical annulus}}.  \jt{Phys. Rev. Fluids}
\bvol{4},  \pg{043501}.

\bibitem[Jiang {\em et~al.\/}(2020)Jiang, Zhu, Wang, Huisman \&
Sun]{JiangSciAdv}
{\sc \au{Jiang, H.}, \au{Zhu, X.}, \au{Wang, D.}, \au{Huisman, S.G.} \&
	\au{Sun, C.}} \yr{2020}  \at{{ Supergravitational turbulent thermal
		convection}}.  \jt{Sci. Adv.}  \bvol{6},  \pg{eabb8676}.

\bibitem[Rouhi {\em et~al.\/}(2021)Rouhi, Lohse, Marusic, Sun \&
Chung]{rouhi2021}
{\sc \au{Rouhi, A.}, \au{Lohse, D.}, \au{Marusic, I.}, \au{Sun, C.} \&
	\au{Chung, D.}} \yr{2021}  \at{{Coriolis effect on centrifugal
		buoyancy-driven convection in a thin cylindrical shell}}.  \jt{J. Fluid
	Mech.}  \bvol{910},  \pg{A32}.

\bibitem[Busse \& Carrigan(1974)]{Busse1}
{\sc \au{Busse, F.H.} \& \au{Carrigan, C.R.}} \yr{1974}  \at{{Convection
		induced by centrifugal buoyancy}}.  \jt{J. Fluid Mech.}  \bvol{62}~(3),
\pg{579–592}.

\bibitem[Azouni {\em et~al.\/}(1985)Azouni, Bolton \& Busse]{Busse2}
{\sc \au{Azouni, M.A.}, \au{Bolton, E.W.} \& \au{Busse, F.H.}} \yr{1985}
\at{{Convection driven by centrifugal bouyancy in a rotating annulus}}.
\jt{Geophys. Astrophys. Fluid Dyn.}  \bvol{34}~(1-4),  \pg{301--317}.

\bibitem[Busse \& Or(1986)]{Busse3}
{\sc \au{Busse, F.H.} \& \au{Or, A.C.}} \yr{1986}  \at{{Convection in a
		rotating cylindrical annulus: thermal Rossby waves}}.  \jt{J. Fluid Mech.}
\bvol{166},  \pg{173–187}.

\bibitem[Busse(1994)]{Busse4}
{\sc \au{Busse, F.H.}} \yr{1994}  \at{{Convection driven zonal flows and
		vortices in the major planets}}.  \jt{Chaos}  \bvol{4}~(2),  \pg{123--134}.

\bibitem[Yano {\em et~al.\/}(2005)Yano, Talagrand \&
Drossart]{Yano_topographic_approx}
{\sc \au{Yano, J.-I.}, \au{Talagrand, O.} \& \au{Drossart, P.}} \yr{2005}
\at{{Deep two-dimensional turbulence: An idealized model for atmospheric jets
		of the giant outer planets}}.  \jt{Geophys. Astrophys. Fluid Dyn.}
\bvol{99}~(2),  \pg{137--150}.

\bibitem[Heimpel {\em et~al.\/}(2005)Heimpel, Aurnou \& Wicht]{Jupiter/Nature}
{\sc \au{Heimpel, M.}, \au{Aurnou, J.} \& \au{Wicht, J.}} \yr{2005}  \at{{
		Simulation of equatorial and high-latitude jets on Jupiter in a deep
		convection model}}.  \jt{Nature}  \bvol{438},  \pg{193--196}.

\bibitem[Porco {\em et~al.\/}(2003)Porco, West, McEwen, Genio, Ingersoll,
Thomas, Squyres, Dones, Murray, Johnson, Burns, Brahic, Neukum, Veverka,
Barbara, T.Denk, Evans, Ferrier, Geissler, Helfenstein, Roatsch, Throop,
Tiscareno \& Vasavada11]{Jupiter/picture}
{\sc \au{Porco, C.C.}, \au{West, R.A.}, \au{McEwen, A.}, \au{Genio, A.D.~Del},
	\au{Ingersoll, A.P.}, \au{Thomas, P.}, \au{Squyres, S.}, \au{Dones, L.},
	\au{Murray, C.D.}, \au{Johnson, T.V.}, \au{Burns, J.A.}, \au{Brahic, A.},
	\au{Neukum, G.}, \au{Veverka, J.}, \au{Barbara, J.M.}, \au{T.Denk},
	\au{Evans, M.}, \au{Ferrier, J.J.}, \au{Geissler, P.}, \au{Helfenstein, P.},
	\au{Roatsch, T.}, \au{Throop, H.}, \au{Tiscareno, M.} \& \au{Vasavada11,
		A.R.}} \yr{2003}  \at{{Cassini Imaging of Jupiter’s Atmosphere, Satellites,
		and Rings}}.  \jt{Science}  \bvol{299},  \pg{1541--1547}.

\bibitem[Rhines(1975)]{rhines_1975}
{\sc \au{Rhines, P.B.}} \yr{1975}  \at{{Waves and turbulence on a beta-plane}}.
\jt{J. Fluid Mech.}  \bvol{69}~(3),  \pg{417–443}.

\bibitem[Sun {\em et~al.\/}(2005)Sun, Ren, Song \& Xia]{sun_ren_song_xia_2005}
{\sc \au{Sun, C.}, \au{Ren, L.-Y.}, \au{Song, H.} \& \au{Xia, K.-Q.}} \yr{2005}
\at{{Heat transport by turbulent {{Rayleigh-B\'enard}} convection in 1 m
		diameter cylindrical cells of widely varying aspect ratio}}.  \jt{J. Fluid
	Mech.}  \bvol{542},  \pg{165–174}.

\bibitem[{van der Poel} {\em et~al.\/}(2011){van der Poel}, Stevens \&
Lohse]{Poel2011PRE}
{\sc \au{{van der Poel}, E.P.}, \au{Stevens, R.J.A.M.} \& \au{Lohse, D.}}
\yr{2011}  \at{{Connecting flow structures and heat flux in turbulent
		{{Rayleigh-B\'enard}} convection}}.  \jt{Phys. Rev. E}  \bvol{84},
\pg{045303}.

\bibitem[Huang {\em et~al.\/}(2013)Huang, Kaczorowski, Ni \& Xia]{Huang2013PRL}
{\sc \au{Huang, S.-D.}, \au{Kaczorowski, M.}, \au{Ni, R.} \& \au{Xia, K.-Q.}}
\yr{2013}  \at{{Confinement-Induced Heat-Transport Enhancement in Turbulent
		Thermal Convection}}.  \jt{Phys. Rev. Lett.}  \bvol{111},  \pg{104501}.

\bibitem[Huang \& Xia(2016)]{huang_xia_2016}
{\sc \au{Huang, S.-D.} \& \au{Xia, K.-Q.}} \yr{2016}  \at{{Effects of geometric
		confinement in quasi-2-D turbulent {{Rayleigh-B\'enard}} convection}}.
\jt{J. Fluid Mech.}  \bvol{794},  \pg{639–654}.

\bibitem[Grossmann {\em et~al.\/}(2016)Grossmann, Lohse \& Sun]{Sun2016Annu}
{\sc \au{Grossmann, S.}, \au{Lohse, D.} \& \au{Sun, C.}} \yr{2016}
\at{{High–Reynolds Number Taylor-Couette Turbulence}}.  \jt{Annu. Rev.
	Fluid Mech.}  \bvol{48}~(1),  \pg{53--80}.

\bibitem[Pitz {\em et~al.\/}(2017{\natexlab{{\em a\/}}})Pitz, Chew, Marxen \&
Hills]{design1}
{\sc \au{Pitz, D.B.}, \au{Chew, J.W.}, \au{Marxen, O.} \& \au{Hills, N.J.}}
\yr{2017{\natexlab{{\em a\/}}}}  \at{{Direct Numerical Simulation of Rotating
		Cavity Flows Using a Spectral Element-Fourier Method}}.  \jt{J. Eng. Gas
	Turb. Power}  \bvol{139}~(7).

\bibitem[King {\em et~al.\/}(2005)King, Wilson \& Owen]{design2}
{\sc \au{King, M.P.}, \au{Wilson, M.} \& \au{Owen, J.M.}} \yr{2005}
\at{{{{Rayleigh-B\'enard}} Convection in Open and Closed Rotating Cavities}}.
\jt{J. Eng. Gas Turb. Power}  \bvol{129}~(2),  \pg{305--311}.

\bibitem[Chalghoum {\em et~al.\/}(2018)Chalghoum, Elaoud, Kanfoudi \&
Akrout]{design3}
{\sc \au{Chalghoum, I.}, \au{Elaoud, S.}, \au{Kanfoudi, H.} \& \au{Akrout, M.}}
\yr{2018}  \at{{The effects of the rotor-stator interaction on unsteady
		pressure pulsation and radial force in a centrifugal pump}}.  \jt{J.
	Hydrodyn.}  \bvol{30}~(4),  \pg{672--681}.

\bibitem[Pitz {\em et~al.\/}(2017{\natexlab{{\em b\/}}})Pitz, Marxen \&
Chew]{pitz_marxen_chew_2017}
{\sc \au{Pitz, D.B.}, \au{Marxen, O.} \& \au{Chew, J.W.}}
\yr{2017{\natexlab{{\em b\/}}}}  \at{{Onset of convection induced by
		centrifugal buoyancy in a rotating cavity}}.  \jt{J. Fluid Mech.}
\bvol{826},  \pg{484–502}.

\bibitem[Verzicco \& Orlandi(1996)]{VERZICCO1996402}
{\sc \au{Verzicco, R.} \& \au{Orlandi, P.}} \yr{1996}  \at{{A Finite-Difference
		Scheme for Three-Dimensional Incompressible Flows in Cylindrical
		Coordinates}}.  \jt{J. Comput. Phys.}  \bvol{123},  \pg{402--414}.

\bibitem[{van der Poel} {\em et~al.\/}(2015{\natexlab{{\em a\/}}}){van der
	Poel}, Ostilla-Mónico, Donners \& Verzicco]{VANDERPOEL201510}
{\sc \au{{van der Poel}, E.P.}, \au{Ostilla-Mónico, R.}, \au{Donners, J.} \&
	\au{Verzicco, R.}} \yr{2015{\natexlab{{\em a\/}}}}  \at{{A pencil distributed
		finite difference code for strongly turbulent wall-bounded flows}}.
\jt{Comput. Fluids}  \bvol{116},  \pg{10--16}.

\bibitem[Zhu {\em et~al.\/}(2018)Zhu, Phillips, Spandan, Donners, Ruetsch,
Romero, Ostilla, Yang, Lohse, Verzicco, Fatica \& Stevens]{Zhu2018comput}
{\sc \au{Zhu, X.}, \au{Phillips, E.}, \au{Spandan, V.}, \au{Donners, J.},
	\au{Ruetsch, G.}, \au{Romero, J.}, \au{Ostilla, R.}, \au{Yang, Y.},
	\au{Lohse, D.}, \au{Verzicco, R.}, \au{Fatica, M.} \& \au{Stevens, R.J.A.M.}}
\yr{2018}  \at{Afid-gpu: A versatile navier–stokes solver for wall-bounded
	turbulent flows on gpu clusters}.  \jt{Comput. Phys. Commun.}  \bvol{229},
\pg{199--210}.

\bibitem[Silano {\em et~al.\/}(2010)Silano, Sreenivasan \&
Verzicco]{Silano2010}
{\sc \au{Silano, G.}, \au{Sreenivasan, K.R.} \& \au{Verzicco, R.}} \yr{2010}
\at{{Numerical simulations of {{Rayleigh-B\'enard}} convection for Prandtl
		numbers between ${10^{ - 1}}$ and ${10^{4}}$ and Rayleigh numbers between
		${10^{5}}$ and ${10^{9}}$}}.  \jt{J. Fluid Mech.}  \bvol{662},
\pg{409–446}.

\bibitem[Courant {\em et~al.\/}(1928)Courant, Friedrichs \& Lewy]{CFL1928}
{\sc \au{Courant, R.}, \au{Friedrichs, K.} \& \au{Lewy, H.}} \yr{1928}
\at{{Über die partiellen Differenzengleichungen der mathematischen}}.
\jt{Physik. Math. Ann.}  \bvol{100},  \pg{32--74}.

\bibitem[Zhang {\em et~al.\/}(2017)Zhang, Zhou \& Sun]{zhang_zhou_sun_2017}
{\sc \au{Zhang, Y.}, \au{Zhou, Q.} \& \au{Sun, C.}} \yr{2017}  \at{{Statistics
		of kinetic and thermal energy dissipation rates in two-dimensional turbulent
		Rayleigh–Bénard convection}}.  \jt{J. Fluid Mech.}  \bvol{814},
\pg{165–184}.

\bibitem[Ostilla {\em et~al.\/}(2013)Ostilla, Stevens, Grossmann, Verzicco \&
Lohse]{ostilla2013}
{\sc \au{Ostilla, R.}, \au{Stevens, R.J.A.M.}, \au{Grossmann, S.},
	\au{Verzicco, R.} \& \au{Lohse, D.}} \yr{2013}  \at{Optimal taylor–couette
	flow: direct numerical simulations}.  \jt{J. Fluid Mech.}  \bvol{719},
\pg{14–46}.

\bibitem[Kunnen {\em et~al.\/}(2016)Kunnen, Ostilla-Mónico, van~der Poel,
Verzicco \& Lohse]{kunnen2016}
{\sc \au{Kunnen, R.P.J.}, \au{Ostilla-Mónico, R.}, \au{van~der Poel, E.P.},
	\au{Verzicco, R.} \& \au{Lohse, D.}} \yr{2016}  \at{{Transition to
		geostrophic convection: the role of the boundary conditions}}.  \jt{J.
	Fluid Mech.}  \bvol{799},  \pg{413–432}.

\bibitem[Blass {\em et~al.\/}(2020)Blass, Zhu, Verzicco, Lohse \&
Stevens]{Blass2020JFM}
{\sc \au{Blass, A.}, \au{Zhu, X.}, \au{Verzicco, R.}, \au{Lohse, D.} \&
	\au{Stevens, R.J.A.M.}} \yr{2020}  \at{{Flow organization and heat transfer
		in turbulent wall sheared thermal convection}}.  \jt{J. Fluid Mech.}
\bvol{897},  \pg{A22}.

\bibitem[{van der Poel} {\em et~al.\/}(2015{\natexlab{{\em b\/}}}){van der
	Poel}, Verzicco, Grossmann \& Lohse]{Poel2015JFM}
{\sc \au{{van der Poel}, E.P.}, \au{Verzicco, R.}, \au{Grossmann, S.} \&
	\au{Lohse, D.}} \yr{2015{\natexlab{{\em b\/}}}}  \at{{Plume emission
		statistics in turbulent {{Rayleigh-B\'enard}} convection}}.  \jt{J. Fluid
	Mech.}  \bvol{772},  \pg{5–15}.

\bibitem[Chong {\em et~al.\/}(2017)Chong, Yang, Huang, Zhong, Stevens,
Verzicco, Lohse \& Xia]{Chong2017PRL}
{\sc \au{Chong, K.L.}, \au{Yang, Y.}, \au{Huang, S.-D.}, \au{Zhong, J.-Q.},
	\au{Stevens, R.J.A.M.}, \au{Verzicco, R.}, \au{Lohse, D.} \& \au{Xia, K.-Q.}}
\yr{2017}  \at{{Confined {{Rayleigh-B\'enard}}, Rotating
		{{Rayleigh-B\'enard}}, and Double Diffusive Convection: A Unifying View on
		Turbulent Transport Enhancement through Coherent Structure Manipulation}}.
\jt{Phys. Rev. Lett.}  \bvol{119},  \pg{064501}.

\bibitem[Jiang {\em et~al.\/}(2018)Jiang, Zhu, Mathai, Verzicco, Lohse \&
Sun]{Jiang2018PRL}
{\sc \au{Jiang, H.}, \au{Zhu, X.}, \au{Mathai, V.}, \au{Verzicco, R.},
	\au{Lohse, D.} \& \au{Sun, C.}} \yr{2018}  \at{{Controlling Heat Transport
		and Flow Structures in Thermal Turbulence Using Ratchet Surfaces}}.
\jt{Phys. Rev. Lett.}  \bvol{120},  \pg{044501}.

\bibitem[Oberbeck(1879)]{Oberbeck1879}
{\sc \au{Oberbeck, A.}} \yr{1879}  \at{{Ueber die Wärmeleitung der
		Flüssigkeiten bei Berücksichtigung der Strömungen infolge von
		Temperaturdifferenzen}}.  \jt{Ann. Phys.-Berlin}  \bvol{243}~(6),
\pg{271--292}.

\bibitem[Wu \& Libchaber(1991)]{Wu1991}
{\sc \au{Wu, X.-Z.} \& \au{Libchaber, A.}} \yr{1991}  \at{{Non-Boussinesq
		effects in free thermal convection}}.  \jt{Phys. Rev. A}  \bvol{43},
\pg{2833--2839}.

\bibitem[Yik {\em et~al.\/}(2020)Yik, Valori \& Weiss]{NOB2020}
{\sc \au{Yik, H.}, \au{Valori, V.} \& \au{Weiss, S.}} \yr{2020}  \at{Turbulent
	{{Rayleigh-B\'enard}} convection under strong non-oberbeck-boussinesq
	conditions}.  \jt{Phys. Rev. Fluids}  \bvol{5},  \pg{103502}.

\bibitem[Zhang {\em et~al.\/}(1997)Zhang, Childress \& Libchaber]{Zhang1997}
{\sc \au{Zhang, J.}, \au{Childress, S.} \& \au{Libchaber, A.}} \yr{1997}
\at{{Non-Boussinesq effect: Thermal convection with broken symmetry}}.
\jt{Phys. Fluids}  \bvol{9}~(4),  \pg{1034--1042}.

\bibitem[Gastine {\em et~al.\/}(2015)Gastine, Wicht \&
Aurnou]{gastine_wicht_aurnou_2015}
{\sc \au{Gastine, T.}, \au{Wicht, J.} \& \au{Aurnou, J.M.}} \yr{2015}
\at{{Turbulent {{Rayleigh-B\'enard}} convection in spherical shells}}.
\jt{J. Fluid Mech.}  \bvol{778},  \pg{721–764}.

\end{thebibliography}

\end{document}